\documentclass[twocolumn,showpacs,aps,prd]{revtex4}

\usepackage{graphicx}
\usepackage{dcolumn}
\usepackage{amsmath}
\usepackage{epsfig}
\usepackage{multirow}

\input babarsym

\def\Kstarzm  {\ensuremath{K^*(892)^0}\xspace}
\def\Kstarzbm {\ensuremath{\Kbar^*(892)^0}\xspace}

\newcommand{\gevcccc}{\ensuremath{{\mathrm{\,Ge\kern -0.1em V^2\!/}c^4}}\xspace}
\newcommand{\BABARPubYear}    {10}
\newcommand{\BABARPubNumber}  {016}

\newcommand{\SLACPubNumber} {14266}
\newcommand{\LANLNumber} {1011.4190}

\def\figurebox#1#2#3{%
    \def\arg{#3}%
    \ifx\arg\empty
    {\hfill\vbox{\hsize#2\hrule\hbox to #2{\vrule\hfill\vbox to #1{\hsize#2\vfill}\vrule}\hrule}\hfill}%
    \else
    {\hfill\epsfbox{#3}\hfill}%
    \fi}

\begin{document}

\preprint{\babar-PUB-\BABARPubYear/\BABARPubNumber} 
\preprint{SLAC-PUB-\SLACPubNumber} 

\begin{flushleft}
\babar-PUB-\BABARPubYear/\BABARPubNumber\\
SLAC-PUB-\SLACPubNumber\\
arXiv:\LANLNumber\ [hep-ex]\\[10mm]
\end{flushleft}

\title{
{\large \bf
Dalitz plot analysis of  {\boldmath$\Ds \to \Kp \Km \pip$}} 
}

%
\author{P.~del~Amo~Sanchez}
\author{J.~P.~Lees}
\author{V.~Poireau}
\author{E.~Prencipe}
\author{V.~Tisserand}
\affiliation{Laboratoire d'Annecy-le-Vieux de Physique des Particules (LAPP), Universit\'e de Savoie, CNRS/IN2P3,  F-74941 Annecy-Le-Vieux, France}
\author{J.~Garra~Tico}
\author{E.~Grauges}
\affiliation{Universitat de Barcelona, Facultat de Fisica, Departament ECM, E-08028 Barcelona, Spain }
\author{M.~Martinelli$^{ab}$}
\author{D.~A.~Milanes}
\author{A.~Palano$^{ab}$ }
\author{M.~Pappagallo$^{ab}$ }
\affiliation{INFN Sezione di Bari$^{a}$; Dipartimento di Fisica, Universit\`a di Bari$^{b}$, I-70126 Bari, Italy }
% ...
\collaboration{The \babar\ Collaboration}
\noaffiliation

\date{\today}

\begin{abstract}
We perform a Dalitz plot analysis of about $100,000$ $\Ds$
decays to $\Kp \Km \pip$ and measure the complex amplitudes of the intermediate resonances which contribute to this decay mode.
We also measure the relative branching fractions of $\Ds \to \Kp \Kp \pim$ and $\Ds \to \Kp \Kp \Km$.
For this analysis we use a 384~${\rm fb}^{-1}$ data sample, 
recorded by the \babar\  detector at the \pep2 asymmetric-energy $e^+e^-$
collider running at center-of-mass energies near 10.58~\gev.
\end{abstract}


\maketitle

\section{Introduction}

Scalar mesons are still a puzzle in light meson spectroscopy. New claims for the existence of broad states close to threshold such as $\kappa(800)$~\cite{Aitala:2002kr} and $f_0(600)$~\cite{Aitala:2000xt}, have reopened discussion about the composition of the ground state
$J^{PC}=0^{++}$ nonet, and about the possibility that states such as the 
$a_0(980)$
or $f_0(980)$ may be 4-quark states, due to their proximity to the
$K \Kbar$ threshold~\cite{Close:2002zu}. This hypothesis 
can be tested only through accurate measurements
of the branching fractions and the couplings to  
different final states. 
It is therefore important to have precise information 
on the structure of the $\pi \pi$ and $K \Kbar$ $\mathcal{S}$-waves. 
In this context, $\Ds$ mesons can shed light on 
the structure of the scalar amplitude coupled to $s \bar s$.
The $\pi \pi$ $\mathcal{S}$-wave has been already extracted from \babar \ data in a Dalitz plot analysis of $\Ds \to \pip \pim \pip$~\cite{:2008tm}. The understanding of the $K \Kbar$ $\mathcal{S}$-wave is also of great importance for the precise measurement of \CP-violation in \Bs oscillations using $\Bs \to \jpsi \phi$~\cite{Stone:2008ak,Xie:2009fs}.

This paper focuses on the study of 
$\Ds$ meson decay to $\Kp\Km\pip$~\cite{conj}.
Dalitz plot analyses of this decay mode have been performed by the E687 and CLEO collaborations using 700 events~\cite{Frabetti:1995sg}, and 14400 events~\cite{:2009tr} respectively. The present analysis is performed using about $100,000$ events.

The decay $\Ds \to \phi \pip$
is frequently used in particle physics as the reference mode for $\Ds$ decay. Previous measurements of this decay mode did not, however, account for the presence of the $K \Kbar$ $\mathcal{S}$-wave underneath the $\phi$ peak.  
Therefore, as part of the present analysis, we obtain a precise measurement
of the branching fraction $\BR(\Ds \to \phi \pip)$ relative to $\BR(\Ds \to K^+ K^- \pi^+)$. 

Singly Cabibbo-suppressed (SCS) and doubly Cabibbo-suppressed (DCS) decays play 
an important role in studies of 
charmed hadron dynamics. The naive expectations for the rates of SCS and DCS decays are of the 
order of $\tan^2 \theta_C$ and 
$\tan^4 \theta_C$, respectively, where $\theta_C$ is the Cabibbo mixing angle. These rates correspond to about 5.3\% and 0.28\% relative to their
Cabibbo-favored (CF) counterpart. 
Due to the limited statistics in past experiments, branching fraction measurements of DCS decays
have been affected by large statistical uncertainties~\cite{Nakamura:2010zzi}. A precise measurement of
$\frac{ \BR(\Ds \to K^+ K^+ \pim)}{ \BR(\Ds \to K^+ K^- \pi^+)}$ has been recently performed by the Belle experiment~\cite{Ko:2009tc}.

In this paper we study the $\Ds$ decay
\begin{equation}
\Ds \to \Kp \Km \pip
\label{eq:eq1}
\end{equation}
\noindent and perform a detailed Dalitz plot analysis.
We then measure the branching ratios of the SCS decay 
\begin{equation}
\Ds \to \Kp \Km \Kp
\label{eq:eq2}
\end{equation}
\noindent and the DCS decay
\begin{equation}
\Ds \to \Kp \Kp \pim
\label{eq:eq3}
\end{equation}
relative to the CF channel (\ref{eq:eq1}).
The paper is organized as follows. Section~\ref{sec:sec_det}  briefly describes the 
\babar\ detector, while Sec.~\ref{sec:sec_ev_sel} gives details of event reconstruction.
Section~\ref{sec:sec_eff} is devoted to the evaluation of the selection efficiency. Section~\ref{sec:sec_pwa} describes a partial
wave analysis of the $\Kp \Km$ system, the evaluation of the $\Ds \to \phi \pip$ branching 
fraction and the $K \Kbar$ $\mathcal{S}$-wave parametrization. Section~\ref{sec:sec_DP_method}
deals with the description of the Dalitz plot analysis method and background description. 
 Results from the Dalitz plot 
analysis of $\Ds \to K^+ K^- \pi^+$ are 
given in Sec.~\ref{sec:sec_DP}. The measurements of the $\Ds$ SCS and DCS branching fractions are described in Sec.~\ref{sec:sec_BR}, while Sec.~\ref{sec:sec_sum} summarizes the results.

\section{The \babar\ Detector and Dataset}
\label{sec:sec_det}
The data sample used in this analysis corresponds to an integrated luminosity
of 384 \invfb recorded 
with the \babar\ detector at the SLAC \pep2\ collider, 
operated at center-of-mass (c.m.) energies near the \Y4S resonance. 
The \babar\ detector is
described in detail elsewhere~\cite{Aubert:2001tu}.
The following is a brief summary of the 
components important to this analysis. 
Charged particle tracks are detected,
and their momenta measured, by a combination of a cylindrical drift chamber (DCH)
and a silicon vertex tracker (SVT), both operating within a
1.5~T solenoidal magnetic field.
Photon energies are measured with a
CsI(Tl) electromagnetic calorimeter (EMC).
Information from a ring-imaging Cherenkov detector (DIRC), and specific energy-loss measurements in the SVT and DCH, are used to identify charged kaon and pion candidates.

\begin{figure*}
\begin{center}
\includegraphics[width=\textwidth]{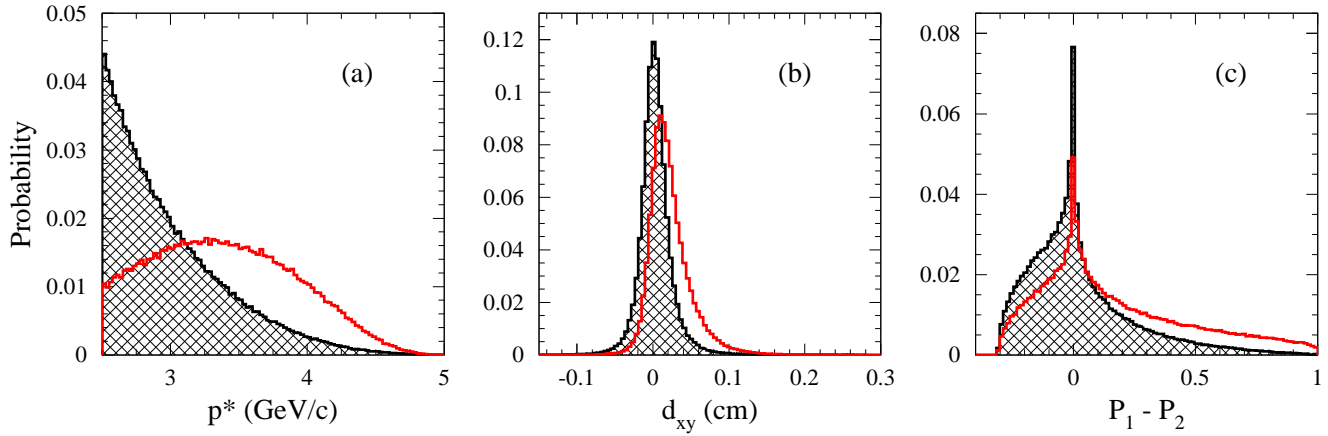}
\caption{Normalized probability distribution functions for signal (solid) and background events (hatched) used in a likelihood-ratio test for the event selection of $D_s^+ \to K^+K^-\pi^+$: (a) the center of mass momentum $p^*$, (b) the signed decay distance $d_{xy}$ and (c) the difference in probability $P_1-P_2$.}
\label{fig:fig1}
\end{center}
\end{figure*}

\section{Event Selection and {\boldmath$\protect\Ds \to \Kp \Km \pip$} Reconstruction}
\label{sec:sec_ev_sel}
Events corresponding to the three-body $\Ds \to \Kp \Km \pip$ decay 
are reconstructed from the data sample having
at least three reconstructed charged tracks with net charge $\pm$ 1. We require that the invariant mass of the $K^+K^-\pi^+$ system lie within the mass interval [1.9-2.05]\gevcc. Particle identification is applied to the three tracks, and the presence of two kaons is required. The efficiency that a kaon is identified is 90\% while the rate that a kaon is misidentified as a pion is 2\%.
The three tracks are required to originate from a common vertex, and the $\chi^2$ fit probability ($P_1$) must be greater than 0.1\%. 
We also perform a separate kinematic fit in which the $\Ds$ mass is constrained to its known value~\cite{Nakamura:2010zzi}. This latter fit will be used only in the Dalitz plot analysis.

In order to help in the discrimination of signal from background, 
an additional fit is performed, constraining the three tracks to originate from the $e^+ e^-$
luminous region (beam spot). 
The $\chi^2$ probability of this fit, labeled as $P_2$, is expected to be
large for most of the background events, when all tracks originate from the luminous region, 
and small for the $\Ds$ signal, due to the measurable flight distance of  the latter.

The decay
\begin{equation} 
D^*_s(2112)^+ \to \Ds \gamma 
\end{equation}
is used to select a subset of event candidates in order to reduce combinatorial 
background. The photon is required to have released an energy of at least 100 \mev into the EMC. We define the variable
\begin{equation}
\Delta m = m(\Kp \Km \pip \gamma) - m(\Kp \Km \pip)
\end{equation}
and require it to be within $\pm 2\sigma_{\Dss}$ with respect to $\Delta m_{\Dss} $ where  $\Delta m_{\Dss}=144.94\pm0.03_{\rm stat}$~\mevcc and $\sigma_{\Dss}=5.53 \pm 0.04_{\rm stat}$~\mevcc are obtained from a Gaussian fit of the $\Delta m$ distribution. 

Each $\Ds$ candidate is characterized by three
variables: the c.m. momentum $p^*$ in the $e^+e^-$ rest frame, the difference in probability $P_1 - P_2$, and the signed decay distance $d_{xy} = \frac{{\mathbf d} \cdot {\mathbf p_{xy}}}{|{\mathbf p_{xy}}|}$ where ${\mathbf d}$ is the vector joining the beam spot to the $\Ds$ decay vertex and ${\mathbf p_{xy}}$ is the projection of the $\Ds$ momentum on the $xy$ plane. These three variables are used to discriminate signal from background events: in fact signal events are expected to be characterized by larger values of $p^*$~\cite{Aubert:2002ue}, due to the jet-like shape of the $e^+e^-\to c \bar c$ events, and larger values of $d_{xy}$ and $P_1-P_2$, due to the measurable flight distance of the $\Ds$ meson.

The distributions of these three variables for signal and background events are determined from data and are shown in Fig.~\ref{fig:fig1}. The background distributions are estimated from events in the $\Ds$ mass-sidebands, while those for the signal region are estimated from the $\Ds$ signal region with sideband subtraction. 
The normalized probability distribution functions (PDFs) are then combined in a likelihood-ratio test. A selection is performed on this variable such that signal to background ratio is maximized. Lower sideband, signal and upper sideband regions are defined between [1.911 - 1.934]~\gevcc, [1.957 - 1.980]~\gevcc and [2.003 - 2.026]~\gevcc, respectively, corresponding to $(-10 \sigma,-6\sigma)$, $(-2 \sigma,2 \sigma)$ and $(6 \sigma, 10 \sigma)$ regions, where $\sigma$ is estimated from the fit of a Gaussian function to the $\Ds$ lineshape.
 
We have examined a number of possible background sources.
A small peak due to the decay $D^{*+} \to \pip \Dz$ where $\Dz \to \Kp \Km$ is observed.
A Gaussian fit to this $\Kp \Km$ spectrum gives $\sigma_{\Dz \to \Kp\Km}=5.4$~\mevcc. For events within 3.5$\sigma_{\Dz \to \Kp\Km}$ of the $D^0$ mass,  
we plot the mass difference $\Delta m(\Kp \Km \pip) = m(\Kp \Km \pip)-m(\Kp \Km)$ and observe a clean $\Dstarp$ signal. We remove events that satisfy $\Delta m(\Kp \Km \pip)< 0.15$~\gevcc.
The surviving events still show a $\Dz \to \Kp \Km$ signal which does not come from this $\Dstarp$ decay. We remove events that satisfy $m(\Kp \Km)>$1.85~\gevcc.

Particle misidentification, in which a pion $\pi_{\rm mis}^+$ is wrongly identified as a kaon, is tested by assigning the pion mass to the \Kp. In this way we identify the background due to the decay $\Dp \to \Km \pip \pip$ which, for the most part, populates the higher mass $\Ds \to \Kp \Km \pip$ sideband. However, this 
cannot be removed without biasing the $\Ds$ Dalitz plot, and so this background is taken into account in the Dalitz plot analysis.

\begin{figure*}
\begin{center}
\includegraphics[width=7.8cm]{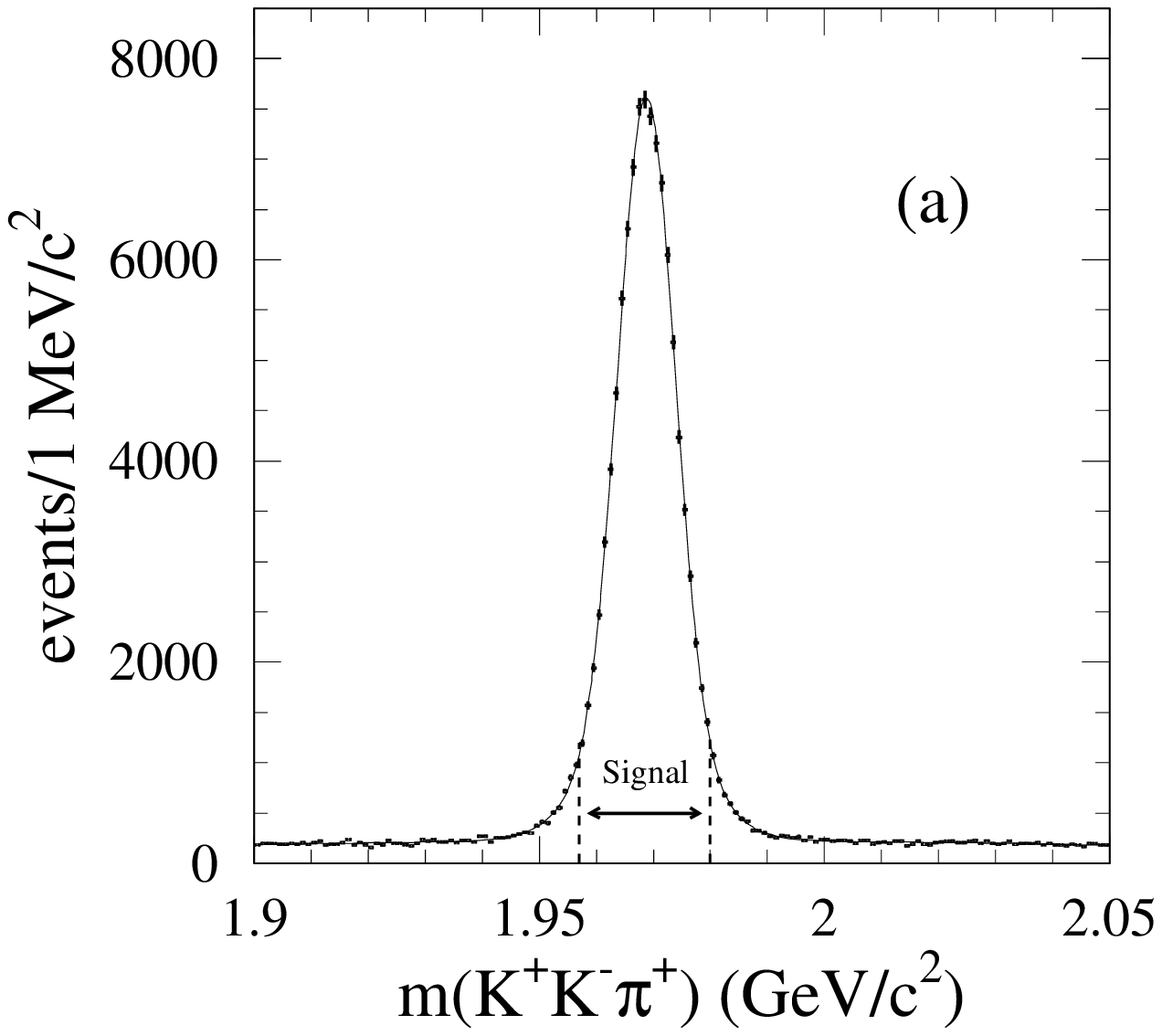}
\includegraphics[width=7.8cm]{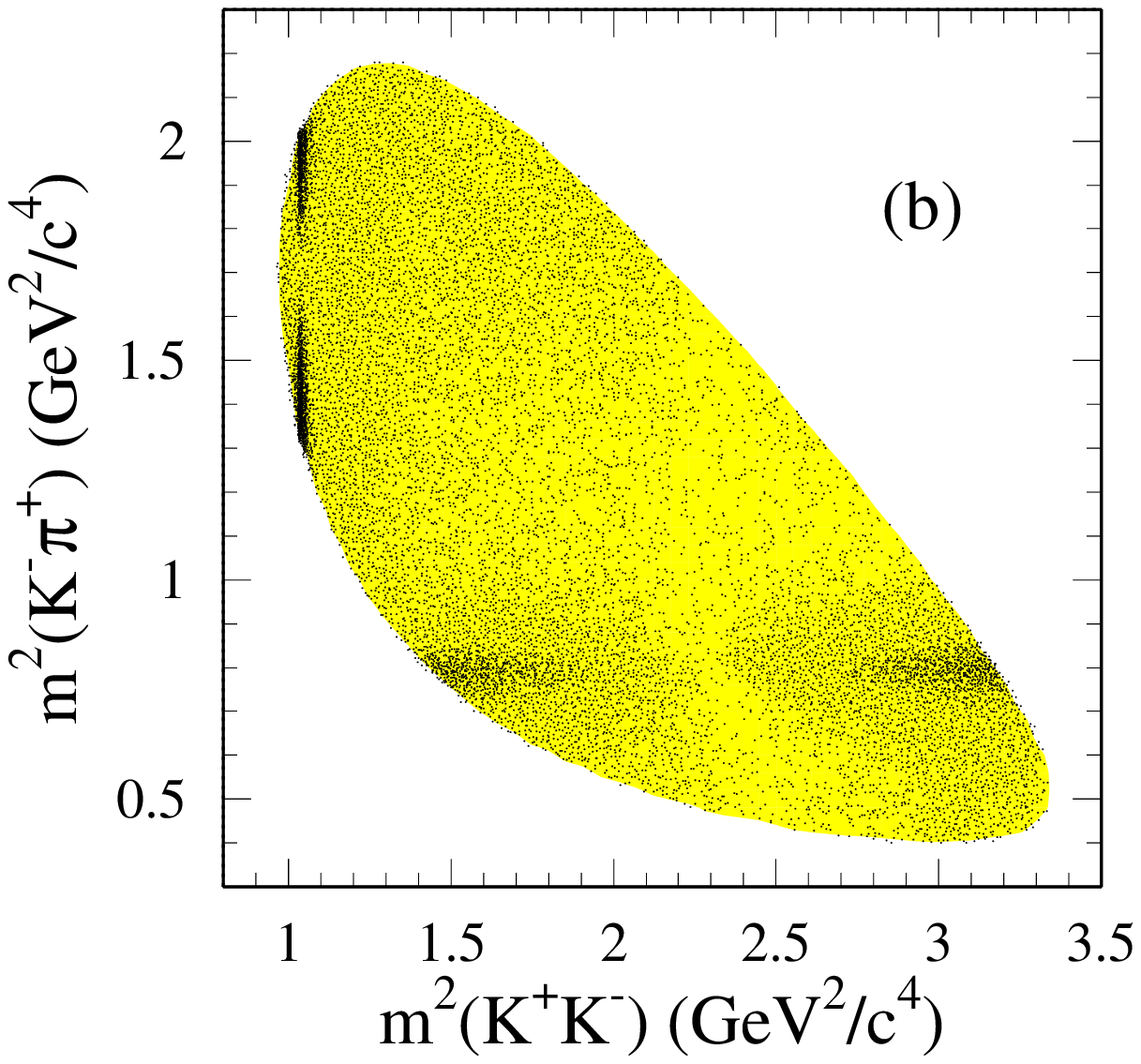}
\caption{(a) $\Kp \Km \pip$ mass distribution for the $\Ds$ analysis sample;
the signal region 
 is as indicated. (b) $\Ds \to \Kp \Km \pip$ Dalitz plot.}
\label{fig:fig2}
\end{center}
\end{figure*}

We also observe
a clean peak in the distribution of the mass difference $m(\Km \pi_{\rm mis}^+ \pip) - m(\Km \pi_{\rm mis}^+)$. 
Combining  $m(K^- \pi_{\rm mis}^+)$ with each of the $\pi^0$ meson candidates in the event, 
we identify this contamination as due to $D^{*+} \to \pi^+ D^0 (\to K^- \pi^+ \pi^0)$
with a missing $\pi^0$. We remove events that satisfy $m(\Km \pip_{\rm mis} \pip)-m(\Km \pip_{\rm mis})<0.15$~\gevcc.
Finally, we remove the $\Ds$ candidates that share one or two daughters with another $\Ds$ candidate; this reduces the number of candidates by 1.8\%, corresponding to 0.9\% of events. We allow there to be two or more non-overlapping multiple candidates in the same event.
The resulting $K^+K^-\pi^+$ mass distribution is shown
in Fig.~\ref{fig:fig2}(a). 
This distribution is fitted with a double-Gaussian function
for the signal, and a linear background. The fit gives a $\Ds$ mass
of  $1968.70 \pm 0.02_{\rm stat}$~\mevcc, $\sigma_1=4.96 \pm 0.06_{\rm stat}$~\mevcc, 
$\sigma_2/\sigma_1=1.91 \pm 0.06_{\rm stat}$ where $\sigma_1$ ($\sigma_2$) is the standard deviation of the first (second) Gaussian, and errors are statistical only.
The fractions of the two Gaussians are $f_{\sigma_1} = 0.80 \pm 0.02$ and $f_{\sigma_2} = 0.20 \pm 0.02$.
The signal region is defined to be within $\pm 2 \sigma_{\Ds}$ of the fitted mass value, where 
$\sigma_{\Ds}=\sqrt {f_{\sigma_1}\sigma_1^2+f_{\sigma_2}\sigma_2^2}=6.1$~\mevcc is the observed mass resolution (the simulated mass resolution is $6$~\mevcc) . The number of signal events in this region (Signal), and the corresponding purity (defined as Signal/(Signal+Background)), are given in Table~\ref{tab:table1}.
\begin{table}[!h]
\caption{Yields and purities for the different $\Ds$ decay modes. Quoted uncertainties are statistical only.}
\begin{center}
\begin{tabular}{crclc}
\hline
$\Ds$ decay mode & \multicolumn{3}{c}{Signal yield} & Purity (\%) \cr
\hline
\hline
$\Kp \Km \pip$ & 96307 & $\pm$ & 369 & 95 \cr
$\Kp \Km \Kp$ & 748 & $\pm$ & 60 & 28 \cr
$\Kp \Kp \pim$ & 356 & $\pm$ & 52 & 23 \cr
\hline
\end{tabular}
\label{tab:table1}
\end{center}
\end{table}

For events in the $\Ds \to \Kp\Km\pip$ signal region,
we obtain the Dalitz plot shown 
in Fig.~\ref{fig:fig2}(b). For this distribution, and for the Dalitz plot analysis (Sec.~\ref{sec:sec_DP_method}), we use the track parameters obtained from the $\Ds$ mass-constrained fit, since this yields a unique Dalitz plot boundary.

In the $\Kp \Km$ threshold region, a strong $\phi(1020)$ signal is observed, 
together with a rather broad structure. 
The $f_0(980)$ and $a_0(980)$ $\mathcal{S}$-wave resonances are, in fact, close to
$K^+ K^-$ 
threshold, and might be expected to contribute in the vicinity of the $\phi(1020)$.
A strong $\Kstarzbm$ signal can also be seen in the $K^-\pi^+$ system, but there is no evidence of structure in the $K^+\pi^+$ mass.

\begin{figure*}
\begin{center}
\includegraphics[width=\textwidth]{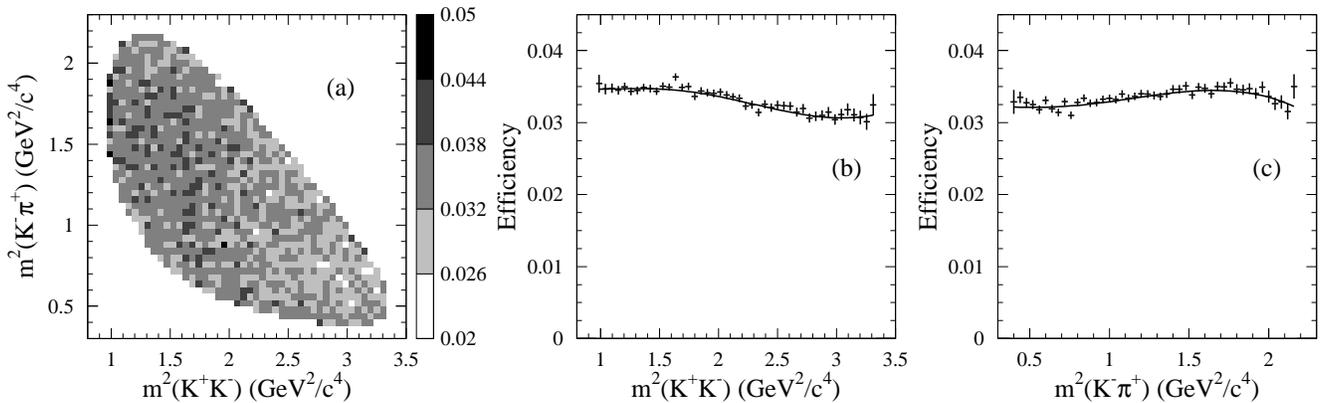}
\caption{(a) Dalitz plot efficiency map; the projection onto (b) the $\Kp\Km$ and (c) the $\Km\pip$ axis.} 
\label{fig:fig3}
\end{center}
\end{figure*}

\section{Efficiency}
\label{sec:sec_eff}
The selection efficiency for each $\Ds$ decay mode analyzed is
determined from a sample of Monte Carlo (MC) events in which the $\Ds$
decay is generated
according to phase space (i.e. such that the Dalitz plot is uniformly
populated). The generated events are passed through a detector
simulation based on the \textsc{Geant4} toolkit~\cite{Agostinelli:2002hh}, and subjected to the same
reconstruction and event selection procedure as that applied to the data. The distribution
of the selected events in each Dalitz plot is then used to
determine the reconstruction efficiency. 
The MC samples used to 
compute these efficiencies consist of 4.2 $\times 10^6$ generated events for $\Ds \to \Kp \Km \pip$ and $\Ds \to \Kp \Kp \pim$, and 0.7 $\times 10^6$ for $\Ds \to K^+ K^- K^+$ .

For $\Ds \to K^+ K^- \pi^+$, 
the efficiency 
distribution is fitted to
a third-order polynomial in two dimensions using the expression:
\begin{eqnarray}
&\eta(x,y) = & a_0 + a_1x^\prime + a_3x^{\prime2} + a_4y^{\prime2} + a_5x^\prime y^\prime \nonumber\\
& & + a_6x^{\prime3} +a_7y^{\prime3}
\end{eqnarray}
\noindent
where $x=m^2(K^+ K^-)$, $y=m^2(K^- \pi^+)$, $x^\prime=x-2$, and $y^\prime=y-1.25$.
Coefficients consistent with zero have been omitted. We obtain a good description of the efficiency with 
$\chi^2/NDF=1133/(1147-7)=0.994$ ($NDF=$ Number of Degrees of Freedom).
The efficiency is found to be almost 
uniform in $K^-\pi^+$ and $K^+K^-$ mass, 
with an average value of $\approx$ 3.3\% (Fig.~\ref{fig:fig3}).

\begin{figure*}
\begin{center}
\includegraphics[width=\textwidth]{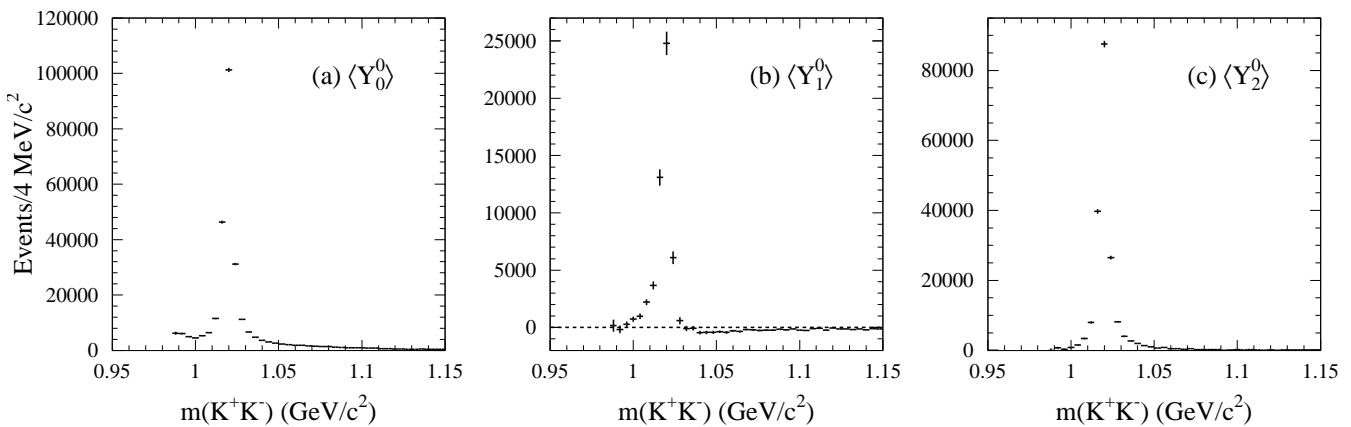}
\caption{$\Kp \Km$ mass spectrum in the threshold region weighted by (a) $Y^0_0$, (b) $Y^0_1$, and (c) $Y^0_2$, corrected for efficiency and phase space, and background-subtracted.}
\label{fig:fig4}
\end{center}
\end{figure*}

\section{Partial Wave Analysis of the {\boldmath$\Kp \Km$} and {\boldmath$\Km \pip$} threshold regions}
\label{sec:sec_pwa}
In the $K^+K^-$ threshold region both $a_0(980)$ and $f_0(980)$ can be present, and both resonances have 
very similar parameters which suffer from large uncertainties.  
In this section we obtain model-independent information on the $\Kp \Km$ $\mathcal{S}$-wave by performing a partial wave analysis in the $\Kp \Km$ threshold region.

Let $N$ be the number of events for a given mass interval $I = [m_{\Kp\Km};m_{\Kp\Km} + {\rm d}m_{\Kp\Km}]$. We write the corresponding angular distribution in terms of the appropriate spherical harmonic functions as
\begin{equation}
\frac{ {\rm d} N}{{\rm d}\cos\theta} = 2\pi\sum_{k=0}^L\left<Y^0_k\right>Y^0_k(\cos\theta),
\label{eq:sph_harmonics}
\end{equation}
\noindent where $L = 2\ell_{\rm max}$, and $\ell_{\rm max}$ is the maximum orbital angular momentum quantum number required to describe the $\Kp\Km$ system
at $m_{\Kp\Km}$ (e.g. $\ell_{\rm max} = 1$ for an $\mathcal{S}$-, $\mathcal{P}$-wave description); $\theta$ is the angle between the $\Kp$ direction in the $\Kp \Km$ rest frame and the prior direction of the $\Kp \Km$ system in the $\Ds$ rest frame. The normalizations are such that
\begin{equation}
\int^1_{-1} Y^0_k(\cos\theta) Y^0_j(\cos\theta) {\rm d}\cos\theta = \frac{\delta_{kj}}{2\pi},
\end{equation}
and it is assumed that the distribution $\frac{{\rm d}N}{{\rm d}\cos\theta}$ has been efficiency-corrected and background-subtracted.

Using this orthogonality condition, the coefficients in
the expansion are obtained from:
\begin{equation}
\left<Y^0_k\right> = \int^1_{-1}Y^0_k(\cos\theta)\frac{{\rm d}N}{{\rm d}\cos \theta} {\rm d}\cos\theta
\end{equation}
\noindent where the integral is given, to a good approximation, 
by $\sum^N_{n=1}Y^0_k(\cos\theta_n)$, where $\theta_n$ is the value of $\theta$ for the $n$-th event.

Figure~\ref{fig:fig4} shows the $\Kp \Km$ mass spectrum up to $1.5 \gevcc$ weighted  
by $Y^0_k(\cos\theta)=\sqrt{(2k+1)/4\pi} \ P_k(\cos\theta)$ for $k=0, 1$ and $2$, where $P_k$ is the Legendre polynomial function of order $k$. These distributions are corrected for efficiency and phase space, and background is subtracted using the $\Ds$ sidebands. 

The number of events $N$ for the mass interval $I$ can be expressed also in terms of the partial-wave amplitudes describing the $K^+K^-$ system. Assuming that only $\mathcal{S}$- and $\mathcal{P}$-wave amplitudes are necessary in this limited region, we can write:
\begin{equation}
\frac{{\rm d}N}{{\rm d}\cos\theta} = 2\pi|\mathcal{S} \, Y^0_0(\cos\theta)+\mathcal{P} \, Y^0_1(\cos\theta)|^2.
\label{eq:pwa_expansion}
\end{equation}

By comparing Eq.~(\ref{eq:sph_harmonics}) and Eq.~(\ref{eq:pwa_expansion})~\cite{Chung:1997qd}, we obtain:
\begin{eqnarray}
\sqrt{4 \pi} \left<Y^0_0 \right> & = & |\mathcal{S}|^2 + |\mathcal{P}|^2 \nonumber \\
\label{eq:sp2} \sqrt{4 \pi} \left<Y^0_1 \right> & = & 2 |\mathcal{S}|  |\mathcal{P}| \cos \phi_{\mathcal{SP}}\\
\sqrt{4 \pi} \left<Y^0_2 \right> & = & \frac{2}{\sqrt 5} |\mathcal{P}|^2 \nonumber
\end{eqnarray}
\noindent
where $\phi_{\mathcal{SP}} = \phi_{\mathcal S} - \phi_{\mathcal P}$ is the phase difference between the $\mathcal{S}$- and $\mathcal{P}$-wave amplitudes. These equations relate the interference 
between
the $\mathcal{S}$-wave ($f_0(980)$, and/or $a_0(980)$, and/or nonresonant) and the $\mathcal{P}$-wave ($\phi(1020)$) to the prominent structure in $\left<Y^0_1 \right>$ (Fig.~\ref{fig:fig4}(b)). The $\left<Y^0_1 \right>$ distribution shows the same behavior as for $\Ds \to K^+K^- e^+ \nu_e$ decay~\cite{Aubert:2008rs}.
The $\left<Y^0_2 \right>$ distribution (Fig.~\ref{fig:fig4}(c)), on the other hand, is consistent with the $\phi(1020)$ lineshape. 

\begin{figure}
\begin{center}
\includegraphics[height=22.15cm]{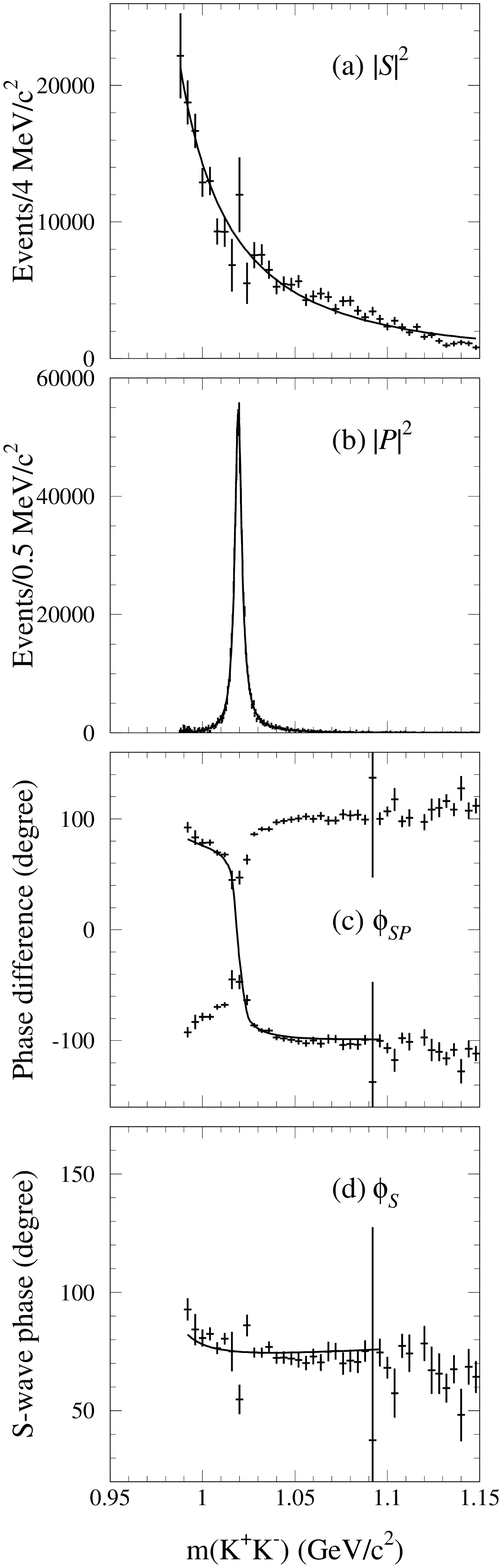}
\caption{Squared (a) $\mathcal{S}$- and (b) $\mathcal{P}$-wave amplitudes; (c) the phase difference $\phi_{\mathcal{SP}}$; (d) $\phi_{\mathcal{S}}$ obtained as explained in the text. The curves result from the fit described in the text.}
\label{fig:fig5}
\end{center}
\end{figure}

The above system of equations can be solved in each interval of $K^+K^-$ invariant mass for $|\mathcal{S}|$, $|\mathcal{P}|$, and $\phi_{\mathcal{SP}}$, and the resulting distributions 
are shown in Fig.~\ref{fig:fig5}. We observe a threshold enhancement in the $\mathcal{S}$-wave (Fig.~\ref{fig:fig5}(a)), and the expected $\phi(1020)$ Breit-Wigner (BW) in the $\mathcal{P}$-wave (Fig.~\ref{fig:fig5}(b)).
We also observe the expected $\mathcal{S}$-$\mathcal{P}$ relative phase motion in the $\phi(1020)$ region (Fig.~\ref{fig:fig5}(c)).

\subsection{{\boldmath$\mathcal{P}$}-wave/{\boldmath$\mathcal{S}$}-wave ratio in the {\boldmath$\phi(1020)$} region}

The decay mode $\Ds \to \phi(1020) \pip$ is used often as the normalizing mode for $\Ds$ decay branching fractions, 
typically by selecting a $\Kp \Km$ invariant mass region around the $\phi(1020)$ peak. 
The observation of a significant $\mathcal{S}$-wave contribution in the threshold region means that this contribution must be taken into account in such a procedure.

In this section we estimate the $\mathcal{P}$-wave/$\mathcal{S}$-wave ratio in an almost model-independent way. 
In fact integrating
the distributions of $\sqrt{4\pi} \, pq^{\prime}\left<Y^0_0\right>$ and $\sqrt{5\pi} \, pq^{\prime}\left<Y^0_2\right>$ (Fig.~\ref{fig:fig4}) in a region around the $\phi(1020)$ peak
yields $\int(|\mathcal{S}|^2+|\mathcal{P}|^2)pq^{\prime}{\rm d}m_{K^+K^-}$ and $\int |\mathcal{P}|^2pq^{\prime}{\rm d}m_{K^+K^-}$ respectively, where $p$ is the $K^+$  momentum in the $K^+K^-$ rest frame, and $q^{\prime}$ is the momentum of the bachelor $\pi^+$ in the $D_s^+$ rest frame.

The $\mathcal{S}$-$\mathcal{P}$ interference contribution integrates to zero, and 
we define the $\mathcal{P}$-wave and $\mathcal{S}$-wave fractions as
\begin{eqnarray}
f_{\mathcal{P}-{\rm wave}} & = &\frac{\int |\mathcal{P}|^2pq^{\prime}{\rm d}m_{K^+K^-}}{\int (|\mathcal{S}|^2+|\mathcal{P}|^2)pq^{\prime}{\rm d}m_{K^+K^-} }\\
f_{\mathcal{S}-{\rm wave}} & = & \frac{\int |\mathcal{S}|^2pq^{\prime}{\rm d}m_{K^+K^-} }{\int (|\mathcal{S}|^2+|\mathcal{P}|^2)pq^{\prime}{\rm d}m_{K^+K^-}} \nonumber\\
& = & 1-f_{\mathcal{P}-{\rm wave}} \, .  
\end{eqnarray}

The experimental mass resolution is estimated by comparing generated and reconstructed MC events, and is 
$\simeq$~0.5~\mevcc at the $\phi$ mass peak.
Table~\ref{tab:table2} gives the resulting $\mathcal{S}$-wave and $\mathcal{P}$-wave fractions computed for three $\Kp \Km$ mass regions. The last column of Table~\ref{tab:table2} shows the measurements of the relative overall rate ($\frac{N}{N_{\rm tot}}$) defined as the number of events in the $\Kp \Km$ mass interval over the number of events in the entire Dalitz plot after efficiency-correction and background-subtraction.

\begin{table}[!htb]
\caption{$\mathcal{S}$-wave and $\mathcal{P}$-wave fractions computed in three $\Kp \Km$ mass ranges around the 
$\phi(1020)$ peak. Errors are statistical only.}
\begin{center}
\begin{tabular}{rclccc}
\hline
\multicolumn{3}{c}{$m_{\Kp \Km}$ (\mevcc)} & $f_{\mathcal{S}-{\rm wave}}$ (\%) & $f_{\mathcal{P}-{\rm wave}}$ (\%) & $\frac{N}{N_{\rm tot}}$ (\%) \\
\hline
\hline
\phantom{11}1019.456 &$\pm$& 5   & 3.5 $\pm$ 1.0 & 96.5 $\pm$ 1.0 & 29.4 $\pm$ 0.2 \\
1019.456 &$\pm$& 10             & 5.6 $\pm$ 0.9 & 94.4 $\pm$ 0.9 & 35.1 $\pm$ 0.2 \\
1019.456 &$\pm$& 15             & 7.9 $\pm$ 0.9 & 92.1 $\pm$ 0.9 & 37.8 $\pm$ 0.2 \\
\hline
\end{tabular}
\end{center}
\label{tab:table2}
\end{table}

\subsection{{\boldmath$\mathcal{S}$}-wave parametrization at the {\boldmath$K^+K^-$} threshold}
\label{sec:sec_pwa_b}

In this section we extract a phenomenological description of the $\mathcal{S}$-wave assuming that it is dominated by the 
$f_0(980)$ resonance while the $\mathcal{P}$-wave is described entirely by the $\phi(1020)$ resonance. 
We also assume that no other contribution is
present in this limited region of the Dalitz plot.
We therefore perform a simultaneous fit of the three distributions shown in Figs.~\ref{fig:fig5}(a),(b), and (c) using the following model:
\begin{equation}
\begin{aligned}
\frac{{\rm d} N_{\mathcal{S}^2}}{{\rm d} m_ {K^+K^-}} = & \, |C_{f_0(980)} A_{f_0(980)}|^2\\
\frac{{\rm d} N_{\mathcal{P}^2}}{{\rm d} m_ {K^+K^-}} = & \, |C_\phi A_\phi|^2\\
\frac{{\rm d} N_{\phi_{\mathcal{S}\mathcal{P}}}}{{\rm d} m_ {K^+K^-}} = & \, arg(A_{f_0(980)}e^{i \delta})-arg(A_\phi)
\end{aligned}
\end{equation}
\noindent
where $C_\phi$, $C_{f_0(980)}$, and $\delta$ are free parameters and
\begin{equation}
A_\phi = \frac{F_r F_D}{m_\phi^2-m^2-im_\phi\Gamma} \times 4 p q
\label{eq:amp_phi}
\end{equation}
\noindent 
is the spin 1 relativistic BW parametrizing the $\phi(1020)$ with $\Gamma$ expressed as:
\begin{equation}
\Gamma = \Gamma_r \left(\frac{p}{p_r}\right)^{2J+1} \left(\frac{M_r}{m}\right)F^2_r.
\label{eq:gamma_phi}
\end{equation}

Here $q$ is the momentum of the bachelor $\pi^+$ in the $K^+K^{-}$ rest frame. The parameters in Eqs.~(\ref{eq:amp_phi}) and (\ref{eq:gamma_phi}) are defined in Sec.~\ref{sec:sec_DP_method} below.

For $A_{f_0(980)}$ we first tried a coupled channel BW (Flatt\'e) amplitude~\cite{Flatte:1972rz}. However we find that this 
parametrization is insensitive to the coupling to the $\pi\pi$ channel. Therefore we empirically parametrize the 
$f_0(980)$ with the following function:
\begin{equation}
A_{f_0(980)} = \frac{1}{m_0^2-m^2-im_0\Gamma_0\rho_{KK}}  
\end{equation}
\noindent
where $\rho_{KK}=2p/m$, and obtain the following parameter values:
\begin{equation}
\begin{aligned}
m_0 = & \ (0.922 \pm 0.003_{\rm stat})\textrm{\gevcc} \\
\Gamma_0 = & \ (0.24 \pm 0.08_{\rm stat}) \ \textrm{GeV}
\end{aligned}
\label{eq:f0_val}
\end{equation}
\noindent The errors are statistical only. The fit results are superimposed on the data in Fig.~\ref{fig:fig5}.

\begin{table}[!tb]
\caption{ $\mathcal{S}$- and $\mathcal{P}$-wave squared amplitudes (in arbitrary units) and $\mathcal{S}$-wave phase. The $\mathcal{S}$-wave phase values, corresponding to the mass 0.988 and 1.116 \gevcc, are missing because the $\left<Y^0_2 \right>$ distribution (Fig.~\ref{fig:fig4}(c)) goes negative or $|\cos\phi_{\mathcal{SP}}|>1$ and so Eqs.~(\ref{eq:sp2}) cannot be solved. Quoted uncertainties are statistical only.}
\begin{center}
\begin{tabular}{crclrclrcl}
\hline
$m_{K^+K^-}$ & \multicolumn{3}{c}{$|\mathcal{S}|^2$} & \multicolumn{3}{c}{$|\mathcal{P}|^2$} & \multicolumn{3}{c}
{$\phi_{\mathcal S}$}\\
(\gevcc) & \multicolumn{3}{c}{(arbitrary units)} & \multicolumn{3}{c}{(arbitrary units)} & \multicolumn{3}{c}{(degrees)}\\
\hline
\hline
0.988 & \phantom{2}22178 &$\pm$& 3120 &  -133  &$\pm$& 2283 & \phantom{100}\\
0.992 & 18760 &$\pm$& 1610 &  2761  &$\pm$& 1313 & 92  &$\pm$& 5 \\
0.996 & 16664 &$\pm$& 1264 &  1043  &$\pm$&  971 & 84  &$\pm$& 7 \\
1     & 12901 &$\pm$& 1058 &  3209  &$\pm$&  882 & 81  &$\pm$& 4 \\
1.004 & 13002 &$\pm$& 1029 &  5901  &$\pm$&  915 & 82  &$\pm$& 3  \\
1.008 & 9300  &$\pm$& 964  &  13484 &$\pm$& 1020 & 76  &$\pm$& 3 \\
1.012 & 9287  &$\pm$& 1117 &  31615 &$\pm$& 1327 & 80  &$\pm$& 2 \\
1.016 & 6829  &$\pm$& 1930 &  157412&$\pm$& 2648 & 75  &$\pm$& 8 \\
1.02  & 11987 &$\pm$& 2734 &  346890&$\pm$& 3794 & 55  &$\pm$& 6 \\
1.024 & 5510  &$\pm$& 1513 &  104892&$\pm$& 2055 & 86  &$\pm$& 5 \\
1.028 & 7565  &$\pm$& 952  &  32239 &$\pm$& 1173 & 75  &$\pm$& 2 \\
1.032 & 7596  &$\pm$& 768  &  15899 &$\pm$&  861 & 74  &$\pm$& 2 \\
1.036 & 6497  &$\pm$& 658  &  10399 &$\pm$&  707 & 77  &$\pm$& 2 \\
1.04  & 5268  &$\pm$& 574  &  7638  &$\pm$&  609 & 72  &$\pm$& 3 \\
1.044 & 5467  &$\pm$& 540  &  5474  &$\pm$&  540 & 72  &$\pm$& 3 \\
1.048 & 5412  &$\pm$& 506  &  4026  &$\pm$&  483 & 72  &$\pm$& 3 \\
1.052 & 5648  &$\pm$& 472  &  2347  &$\pm$&  423 & 71  &$\pm$& 3 \\
1.056 & 4288  &$\pm$& 442  &  3056  &$\pm$&  421 & 70  &$\pm$& 3 \\
1.06  & 4548  &$\pm$& 429  &  1992  &$\pm$&  384 & 73  &$\pm$& 3 \\
1.064 & 4755  &$\pm$& 425  &  1673  &$\pm$&  374 & 70  &$\pm$& 4 \\
1.068 & 4508  &$\pm$& 393  &  1074  &$\pm$&  334 & 75  &$\pm$& 4 \\
1.072 & 3619  &$\pm$& 373  &  1805  &$\pm$&  345 & 75  &$\pm$& 4 \\
1.076 & 4189  &$\pm$& 368  &   840  &$\pm$&  312 & 70  &$\pm$& 5  \\
1.08  & 4215  &$\pm$& 367  &   770  &$\pm$&  297 & 71  &$\pm$& 5 \\
1.084 & 3508  &$\pm$& 345  &   866  &$\pm$&  294 & 71  &$\pm$& 5 \\
1.088 & 3026  &$\pm$& 322  &   929  &$\pm$&  285 & 75  &$\pm$& 4 \\
1.092 & 3456  &$\pm$& 309  &    79  &$\pm$&  240 & 37  &$\pm$& 90 \\
1.096 & 2903  &$\pm$& 300  &   488  &$\pm$&  256 & 75  &$\pm$& 6 \\
1.1   & 2335  &$\pm$& 282  &   885  &$\pm$&  248 & 68  &$\pm$& 5 \\
1.104 & 2761  &$\pm$& 284  &   341  &$\pm$&  231 & 57  &$\pm$& 10 \\
1.108 & 2293  &$\pm$& 273  &   602  &$\pm$&  231 & 77  &$\pm$& 5 \\
1.112 & 1913  &$\pm$& 238  &   269  &$\pm$&  186 & 74  &$\pm$& 8 \\
1.116 & 2325  &$\pm$& 252  &    57  &$\pm$&  198 &     &     & \\
1.12  & 1596  &$\pm$& 228  &   308  &$\pm$&  194 & 78  &$\pm$& 7 \\
1.124 & 1707  &$\pm$& 224  &   233  &$\pm$&  188 & 67  &$\pm$& 10 \\
1.128 & 1292  &$\pm$& 207  &   270  &$\pm$&  176 & 66  &$\pm$& 9 \\
1.132 & 969   &$\pm$& 197  &   586  &$\pm$&  172 & 60  &$\pm$& 6 \\
1.136 & 1092  &$\pm$& 196  &   553  &$\pm$&  170 & 67  &$\pm$& 6 \\
1.14  & 1180  &$\pm$& 193  &   316  &$\pm$&  167 & 48  &$\pm$& 11 \\
1.144 & 1107  &$\pm$& 187  &   354  &$\pm$&  170 & 68  &$\pm$& 8\\
1.148 & 818   &$\pm$& 178  &   521  &$\pm$&  164 & 64  &$\pm$& 7 \\
\hline
\end{tabular}
\end{center}

\label{tab:table3}
\end{table}

\begin{figure*}
\begin{center}
\includegraphics[width=15.3cm]{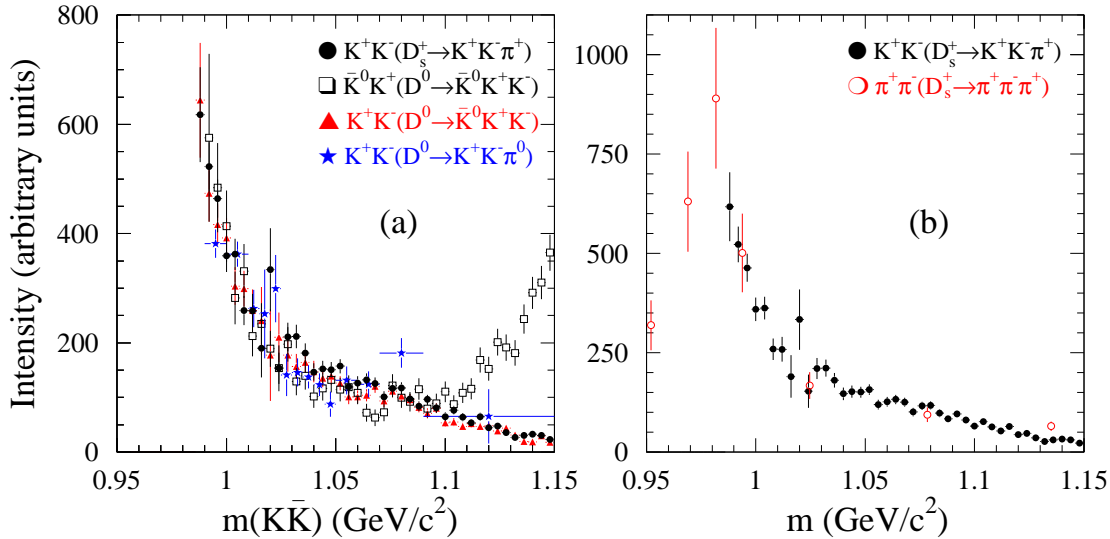}
\caption{(a) Comparison between $K \Kbar$ $\mathcal{S}$-wave intensities from different charmed meson Dalitz plot analyses. (b) Comparison of the $K \Kbar$ $\mathcal{S}$-wave intensity from $\Ds \to \Kp \Km \pip$ with the $\pip \pim$ $\mathcal{S}$-wave intensity from $\Ds \to \pip \pim \pip$.}
\label{fig:fig6}
\end{center}
\end{figure*}

In Fig.~\ref{fig:fig5}(c), the $\mathcal{S}$-$\mathcal{P}$ phase difference is plotted twice because of the sign ambiguity associated with the value of $ \phi_{\mathcal{SP}}$ extracted from $\cos \phi_{\mathcal{SP}}$. We can extract the mass-dependent $f_0(980)$ phase by adding the mass-dependent $\phi(1020)$ BW phase to the $\phi_{\mathcal{SP}}$ distributions of Fig.~\ref{fig:fig5}(c). Since the $K^+ K^-$ mass region is significantly above the $f_0(980)$ central mass value of Eq.~(\ref{eq:f0_val}), we expect that the $\mathcal{S}$-wave phase will be moving much more slowly in this region than in the $\phi(1020)$ region. Consequently, we resolve the phase ambiguity of Fig.~\ref{fig:fig5}(c) by choosing as the physical solution the one which decreases rapidly in the $\phi(1020)$ peak region, since this reflects the rapid forward BW phase motion associated with a narrow resonance. The result is shown in Fig.~\ref{fig:fig5}(d), where we see that the $\mathcal{S}$-wave phase is roughly constant, as would be expected for the tail of a resonance. The slight decrease observed with increasing mass might be due to higher mass contributions to the $\mathcal{S}$-wave amplitude. The values of $|\mathcal{S}|^2$ (arbitrary units) and phase values are reported in Table~\ref{tab:table3}, together with the corresponding values of $|\mathcal{P}|^2$.

In Fig.~\ref{fig:fig6}(a) we compare  the $\mathcal{S}$-wave profile from this analysis with the $\mathcal{S}$-wave intensity values extracted from Dalitz plot analyses of $\Dz \to \Kzb \Kp \Km$~\cite{Aubert:2005sm}
and $\Dz \to \Kp \Km \piz$~\cite{Aubert:2007dc}.
The four distributions are normalized in the region from 
threshold up to 1.05~\gevcc. We observe substantial agreement. 
As the $a_0(980)$ and $f_0(980)$ mesons couple mainly to the $u \bar u/d \bar d$ and $s \bar s$ systems respectively, the former is favoured in $\Dz \to \Kzb \Kp \Km$ and the latter in $\Ds \to \Kp \Km \pip$. Both resonances can contribute in $\Dz \to \Kp \Km \piz$. We conclude that the $\mathcal{S}$-wave projections in the $K \Kbar$ system for both resonances are consistent in shape.
It has been suggested that this feature supports the hypothesis that the $a_0(980)$ and $f_0(980)$ are 4-quark states~\cite{Maiani:2007iw}. We also compare the $\mathcal{S}$-wave profile from this analysis with the $\pip \pim$ $\mathcal{S}$-wave profile extracted from \babar \ data in a Dalitz plot analysis of $\Ds \to \pip \pim \pip$~\cite{:2008tm} (Fig.~\ref{fig:fig6}(b)). The observed agreement supports the argument that only the $f_0(980)$ is present in this limited mass region.

\subsection{Study of the {\boldmath$K^-\pi^+$} {\boldmath$\mathcal{S}$}-wave at threshold}
\label{sec:kpi_swave}
\begin{figure*}
\begin{center}
\includegraphics[width=\textwidth]{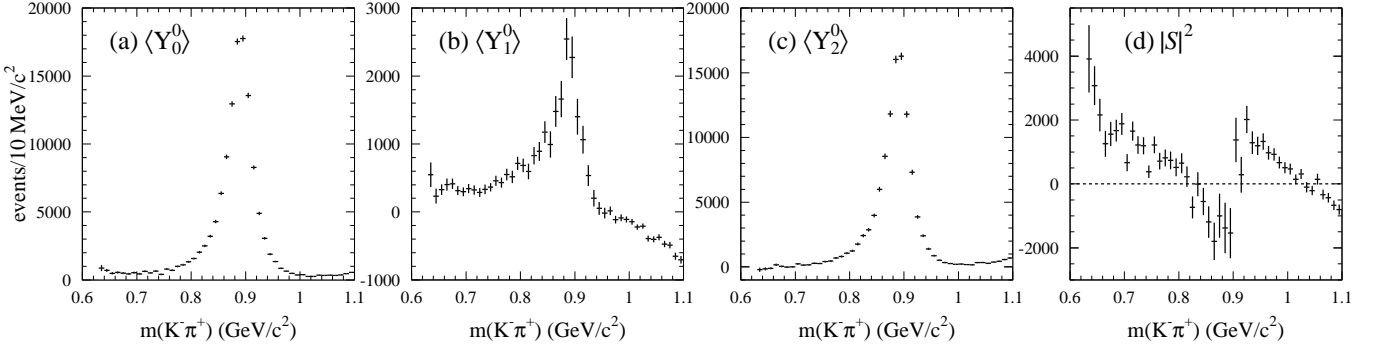}
\caption{$\Km \pip$ mass spectrum in the threshold region weighted by (a) $Y^0_0 $, (b) $Y^0_1$ and (c) $Y^0_2$, corrected for efficiency, phase space, and background-subtracted. (d) The $K^-\pi^+$ mass dependence of $|\mathcal{S}|^2$.}
\label{fig:fig7}
\end{center}
\end{figure*}

We perform a model-independent analysis, similar to that described in the previous sections, to extract the  $K \pi$ $\mathcal{S}$-wave 
behavior as a function of mass
in the threshold region up to $1.1 \gevcc$. Figure~\ref{fig:fig7} shows the $\Km \pip$ mass spectrum in this region, weighted  
by $Y^0_k(\cos\theta)=\sqrt{(2k+1)/4\pi} P_k(\cos\theta)$, with $k=0, 1$ and $2$,  corrected for efficiency, phase space,  
and with background from the $\Ds$ sidebands subtracted;  $\theta$ is the angle between the $K^-$ direction in the $K^- \pi^+$ rest frame and the prior direction of the $K^- \pi^+$ system in the $D^+_s$ rest frame. We observe that  $\left<Y^0_0 \right>$ and $\left<Y^0_2 \right>$ show strong $\Kstarzbm$ resonance signals, and that the $\left<Y^0_1 \right>$ moment shows evidence for  $\mathcal{S}$-$\mathcal{P}$ interference. 

We use Eqs.~(\ref{eq:sp2})
to solve for $|\mathcal{S}|$ and $|\mathcal{P}|$. The result
for the $\mathcal{S}$-wave is shown in Fig.~\ref{fig:fig7}(d). 
We observe a small $\mathcal{S}$-wave contribution which does not allow us to measure the expected 
phase motion relative to that of the $\Kstarzbm$ resonance. Indeed, the fact that $|\mathcal{S}|^2$ goes negative indicates that a model including only $\mathcal{S}$- and $\mathcal{P}$-wave components is not sufficient to describe the $K^-\pi^+$ system.

\section{Dalitz Plot formalism}
\label{sec:sec_DP_method}
An unbinned maximum likelihood fit is performed
in which the distribution of events 
in the Dalitz plot is used to determine the relative amplitudes and phases 
of intermediate resonant and nonresonant states.
 
The likelihood function is written as:
\begin{eqnarray}
\mathcal{L} =  \prod_{n=1}^N&\bigg[&f_{\rm sig} \cdot \eta(x,y)\frac{\sum_{i,j} c_i c_j^* A_i(x,y) A_j^*(x,y)}{\sum_{i,j} c_i c_j^* I_{A_i A_j^*}} + \nonumber\\
& &(1-f_{\rm sig})\frac{\sum_{i} k_iB_i(x,y)}{\sum_{i} k_iI_{B_i}}\bigg]
\end{eqnarray}
\noindent where:
\begin{itemize}
\item $N$ is the number of events in the signal region;
\item $x=m^2(K^+ K^-)$ and $y=m^2(K^- \pi^+)$
\item $f_{\rm sig}$ is the fraction of signal as a function of the $\Kp \Km \pip$ invariant mass,  obtained from the fit to the $\Kp \Km \pip$ mass spectrum (Fig.~\ref{fig:fig2}(a));
\item $\eta(x,y)$ is the efficiency, parametrized by a $3^{\rm rd}$ order polynomial (Sec.~\ref{sec:sec_eff});
\item the $A_i(x,y)$ describe the complex signal amplitude contributions;
\item the $B_i(x,y)$ describe the background probability density function contributions;
\item $k_i$ is the magnitude of the $i$-th component for the background. The $k_i$ parameters are obtained by fitting the sideband regions;
\item $I_{A_i A_j^*}=\int A_i (x,y)A_j^*(x,y) \eta(x,y) {\rm d}x{\rm d}y$ and 
$I_{B_i}~=~\int B_i(x,y) {\rm d}x{\rm d}y$ are normalization
 integrals. Numerical integration is performed by means of Gaussian quadrature~\cite{cern};
\item $c_i$ is the complex amplitude of the $i$-th component for the signal. The $c_i$ parameters are allowed to vary during the fit process.
\end{itemize}

The phase of each amplitude (i.e.\ the phase of the corresponding $c_i$) is measured with respect to 
the $\Kp \Kstarzbm$ amplitude. 
Following the method described in Ref.~\cite{Asner:2003gh}, 
each amplitude $A_i(x,y)$ is represented by the product of a complex BW and a real angular term $T$ depending on the solid angle $\Omega$:
\begin{equation}
A(x,y) = BW(m) \times T (\Omega). 
\end{equation}
For a $D_s$ meson decaying into three pseudo-scalar mesons via an intermediate resonance $r$ ($D_s \to r C, r \to AB$),  
$BW(M_{AB})$ is written as a relativistic BW:
\begin{equation}
BW(M_{AB}) = \frac{F_r F_D}{M_r^2 - M_{AB}^2 - i \Gamma_{AB}M_r}
\end{equation}
\noindent where $\Gamma_{AB}$ is a function of the invariant mass of system $AB$ ($M_{AB}$), the momentum $p_{AB}$ of either daughter in the $AB$ rest frame, the spin $J$ of the resonance and the mass $M_r$ and the width $\Gamma_r$ of the resonance. The explicit expression is:
\begin{equation}
\Gamma_{AB} = \Gamma_r \left(\frac{p_{AB}}{p_r}\right)^{2J+1} \left(\frac{M_r}{M_{AB}}\right)F^2_r
\label{eq:gamma}
\end{equation}
\begin{equation}
p_{AB} = \frac{\sqrt{\left(M_{AB}^2-M_A^2-M_B^2\right)^2-4M_A^2M_B^2}}{2M_{AB}}.
\label{eq:pAB}
\end{equation}

The form factors $F_r$ and $F_D$ attempt to model the underlying quark structure of the parent particle and the intermediate
resonances. We use the Blatt-Weisskopf penetration factors~\cite{blatt} (Table~\ref{tab:table4}), that depend on a single parameter $R$ representing the meson ``radius''. We assume $R_{\Ds}=3 \gev^{-1}$ for the $D_s$ and $R_r=1.5 \gev^{-1}$ for the intermediate resonances; $q_{AB}$ is the momentum of the bachelor $C$ in the $AB$ rest frame:
\begin{equation}
q_{AB} = \frac{\sqrt{\left(M_{D_s}^2+M_C^2-M_{AB}^2\right)^2-4M_{D_s}^2M_C^2}}{2M_{AB}}.
\label{eq:qAB}
\end{equation}
$p_r$ and $q_r$ are the values of $p_{AB}$ and $q_{AB}$ when $m_{AB}=m_r$. 

\begin{figure*}
\begin{center}
\includegraphics[width=\textwidth]{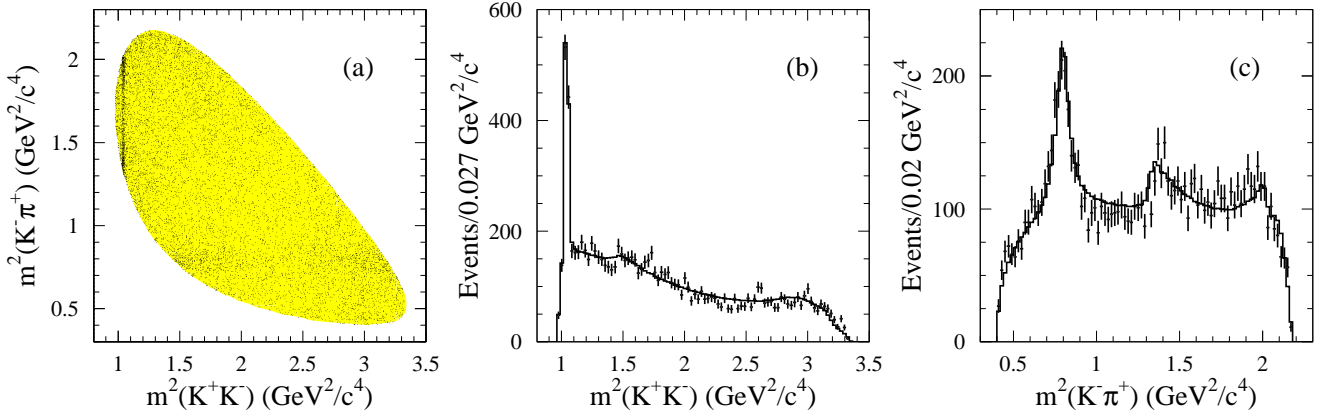}
\caption{(a) Dalitz plot of sideband regions projected onto (b) the $\Kp\Km$ and (c) the $\Km\pip$ axis.} 
\label{fig:fig8}
\end{center}
\end{figure*}

\begin{table}[!htb]
\caption{Summary of the Blatt-Weisskopf penetration form factors. $q_r$ and $p_r$ are the momenta of the decay particles in the parent rest frame.}
\begin{center}
\begin{tabular}{ccc}
\hline 
Spin & $F_r$ & $F_D$\\
\hline \hline
0 & $1$ & $1$\\
&&\\
1 & $\frac{\sqrt{1+(R_r p_r)^2}}{\sqrt{1+(R_r p_{AB})^2}}$ & $\frac{\sqrt{1+(R_{\Ds} q_r)^2}}{\sqrt{1+(R_{\Ds} q_{AB})^2}}$\\
&&\\
2 & $\frac{\sqrt{9+3(R_r p_r)^2+(R_r p_r)^4}}{\sqrt{9+3(R_r p_{AB})^2+(R_r p_{AB})^4}}$ & $\frac{\sqrt{9+3(R_{\Ds} q_r)^2+(R_{\Ds} q_r)^4}}{\sqrt{9+3(R_{\Ds} q_{AB})^2+(R_{\Ds} q_{AB})^4}}$\\
\hline
\end{tabular}
\label{tab:table4}
\end{center}
\end{table}

The angular 
terms $T (\Omega)$ are described by the
following expressions:
\begin{equation}
\begin{aligned}
\textrm{Spin 0}: T(\Omega) = & 1\\
\textrm{Spin 1}: T(\Omega) = & M^2_{BC}-M^2_{AC} \\
& -\frac{(M^2_{D_s}-M^2_C)(M^2_B-M^2_A)}{M_{AB}^2} \\
\textrm{Spin 2}: T(\Omega) = & a_1^2 - \frac{1}{3}a_2 a_3
\end{aligned}
\end{equation}
\noindent where:
\begin{equation}
\begin{aligned}
a_1 = & M^2_{BC}-M^2_{AC}+\frac{(M^2_{D_s}-M^2_C)(M^2_A-M^2_B)}{M_{AB}^2}\\
a_2 = & M^2_{AB}-2M^2_{D_s}-2M^2_C+\frac{(M^2_{D_s}-M^2_C)^2}{M_{AB}^2} \\
a_3 = & M^2_{AB}-2M^2_A-2M^2_B+\frac{(M^2_A-M^2_B)^2}{M^2_{AB}}.
\end{aligned}
\end{equation}

Resonances are included in sequence, starting from those immediately visible 
in the Dalitz plot projections. All allowed resonances from Ref.~\cite{Nakamura:2010zzi} have been tried, and we reject those with amplitudes consistent with zero. The goodness of fit is tested by an adaptive binning $\chi^2$.

The efficiency-corrected fractional contribution due to the resonant or nonresonant contribution $i$ is defined as follows:
\begin{equation}
f_i = \frac {|c_i|^2 \int |A_i(x,y)|^2 {\rm d}x {\rm d}y}
{\int |\sum_j c_j A_j(x,y)|^2 {\rm d}x {\rm d}y}.
\end{equation}
The $f_i$ do not necessarily add to 1 because of interference effects. We also define the interference fit fraction between the resonant or nonresonant contributions $k$ and $l$  as:
\begin{equation}
f_{kl} = \frac {2 \int \Re[c_kc_l^* A_k(x,y) A_l^*(x,y)] {\rm{d}}x {\rm{d}}y}
{\int |\sum_j c_j A_j(x,y)|^2 {\rm d}x {\rm d}y}.
\end{equation}
Note that $f_{kk}=2f_k$. The error on each $f_i$ and $f_{kl}$ is evaluated by propagating the full covariance matrix obtained from the fit. 

\subsection{Background parametrization}

To parametrize the $\Ds$ background, we use the $\Ds$ sideband regions. 
An unbinned maximum likelihood fit is performed using the function:
\begin{equation}
\mathcal{L} = \prod_{n = 1}^{N_B} \left[ \frac{\sum_{i} k_iB_i}{\sum_{i} k_i I_{B_i}} \right]
\end{equation}
\noindent where $N_B$ is the number of sideband events, the $k_i$ parameters are real coefficients floated in the fit, and the 
$B_i$ parameters represent Breit-Wigner functions that are summed incoherently.

The Dalitz plot for the two sidebands shows the presence of $\phi(1020)$ and $\Kstarzbm$ (Fig.~\ref{fig:fig8}). There are further structures not clearly associated with known resonances and due to reflections of other final states. Since they do not have definite spin, we parametrize the background using an incoherent sum of $\mathcal{S}$-wave Breit-Wigner shapes.

\section{Dalitz plot analysis of {\boldmath{$\protect \Ds \to \Kp \Km \pip$}}}
\label{sec:sec_DP}
Using the method described in Sec.~\ref{sec:sec_DP_method}, we perform an unbinned maximum likelihood fit to the $\Ds \to \Kp \Km \pip$
decay channel.
The fit is performed in steps, by adding resonances one after the other. Most of the masses and widths of 
the resonances are taken from Ref.~\cite{Nakamura:2010zzi}. 
For the $f_0(980)$ we use the phenomenological model described in Sec.~\ref{sec:sec_pwa_b}.
The $\Kstarzbm$ amplitude is chosen as the reference amplitude.  

\begin{table*}
\caption{Results from the $\Ds \to K^+ K^- \pi^+$ Dalitz plot analysis. The table gives fit fractions, amplitudes and phases from the best fit. Quoted uncertainties are statistical and systematic, respectively.}
\begin{center}
\begin{tabular}{l r@{}c@{}l r@{}c@{}l r@{}c@{}l}
\hline
Decay mode                   & \multicolumn{3}{c}{Decay fraction (\%)}& \multicolumn{3}{c}{Amplitude}          & \multicolumn{3}{c}{Phase (radians)}            \\
\hline
\hline
$\Kstarzbm K^+$\phantom{{\LARGE I}} & $47.9 \, \pm \, $& $0.5$&$\, \pm \, 0.5$    & \multicolumn{3}{c}{$1. \, ({\rm Fixed})$}               & \multicolumn{3}{c}{$0. \, ({\rm Fixed})$}      \\
$\phi(1020) \, \pi^+$            & $41.4 \, \pm \, $& $0.8$&$\, \pm \, 0.5$    & $1.15 \, \pm \, $&$ 0.01$&$\, \pm \,0.26$  &  $ 2.89 \,  \pm \, $&$ 0.02$&$\, \pm \, 0.04$   \\
$f_0(980) \, \pi^+$              & $16.4 \,   \pm \, $& $0.7$&$\, \pm \, 2.0$  & $2.67  \, \pm \,$ &$ 0.05$  &$\, \pm \, 0.20$   &  $ 1.56 \, \pm \, $&$ 0.02$&$\, \pm \, 0.09$\\
$\Kbar^*_0(1430)^0 K^+$     & $2.4 \,  \pm \, $& $0.3$&$\, \pm \, 1.0$    & $1.14 \, \pm \, $ &$ 0.06$ &$\, \pm \, 0.36$  &  $2.55  \, \pm \, $&$ 0.05$&$\, \pm \, 0.22$  \\
$f_0(1710) \, \pi^+$             & $1.1  \, \pm \, $& $0.1$&$\, \pm \, 0.1$    & $0.65  \, \pm \, $&$ 0.02$ &$\, \pm \, 0.06$  &  $ 1.36 \, \pm \, $&$ 0.05$&$\, \pm \, 0.20$ \\
$f_0(1370) \, \pi^+$             & $1.1  \, \pm \, $& $0.1$&$\, \pm \, 0.2$    & $0.46  \, \pm \, $&$ 0.03$ &$\, \pm \, 0.09$  &  $ -0.45 \, \pm \, $&$ 0.11$&$\, \pm \, 0.52$\\
\hline
Sum                          & $110.2 \, \pm \, $ & $0.6$&$\, \pm \, 2.0$    &                    &&&&&\\
$\chi^2/NDF$                 & \multicolumn{4}{c}{$2843/(2305-14)=1.24$}\\
\hline
\end{tabular}
\label{tab:table5}
\end{center}
\end{table*}

The decay fractions, amplitudes, and relative phase values for the best fit  obtained, are summarized in Table~\ref{tab:table5} where the first error is statistical, and the second is systematic. The interference fractions are quoted in Table~\ref{tab:table6} where the error is statistical only. We observe the following features.
\begin{itemize}
\item{} The decay is dominated by the $\Kstarzbm \Kp$ and $\phi(1020) \pi^+$ amplitudes. 
\item{}
The fit quality is substantially improved by leaving the $\Kstarzbm$
parameters free in the fit. The fitted parameters are:
\begin{equation}
\begin{aligned}
m_{\Kstarzbm} = & \, (895.6 \pm 0.2_{\rm stat}  \pm 0.3_{\rm sys})  \mevcc \\ 
\Gamma_{\Kstarzbm} = & \, (45.1 \pm 0.4_{\rm stat}  \pm 0.4_{\rm sys}) \mev 
\label{eq:m_g_892}
\end{aligned}
\end{equation}

\begin{figure*}
\begin{center}
\includegraphics[width=14.1cm]{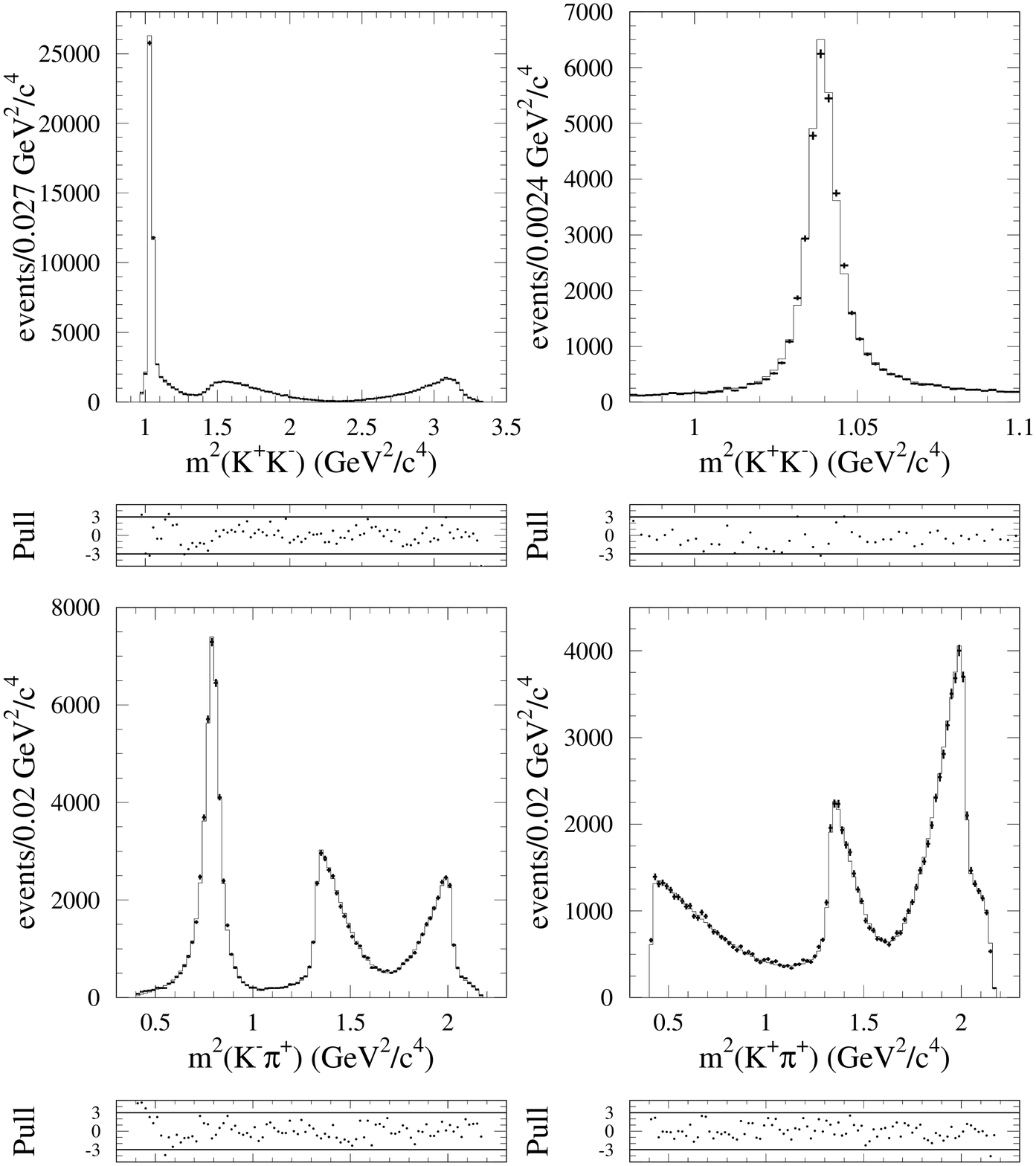}
\caption{$\Ds \to K^+ K^- \pi^+$: Dalitz plot projections from the best fit. The data are represented by points with error bars, the fit results by the histograms.}
\label{fig:fig9}
\end{center}
\end{figure*}

\begin{table*}
\caption{Fit fractions matrix of the best fit. The diagonal elements $f_i$ correspond to the decay fractions in Table~\ref{tab:table5}. The off-diagonal elements give the fit fractions of the interference $f_{kl}$. The null values originate from the fact that any $\mathcal{S}$-$\mathcal{P}$ interference contribution integrates to zero. Quoted uncertainties are statitistical only.}
\begin{center}
\begin{tabular}{l|c rcl rcl rcl rcl rcl}
\hline
$f_{kl}$ (\%)\phantom{{\LARGE I}} & $\Kstarzbm K^+$ & \multicolumn{3}{c}{$\phi(1020) \, \pi^+$} & \multicolumn{3}{c}{$f_0(980) \, \pi^+$} & \multicolumn{3}{c}{$\Kbar^*_0(1430)^0K^+$} & \multicolumn{3}{c}{$f_0(1710) \, \pi^+$} &  \multicolumn{3}{c}{$f_0(1370) \, \pi^+$}\\
\hline
$\Kstarzbm K^+$\phantom{{\LARGE I}} & 47.9 $\pm$ 0.5 & -4.36&$\pm$&0.03  & -2.4&$\pm$&0.2  & \multicolumn{3}{c}{0.}           & -0.06&$\pm$&0.03 & 0.08&$\pm$&0.08\\
 $\phi(1020) \, \pi^+$                  &              & 41.4 &$\pm$&0.8 &    \multicolumn{3}{c}{0.}        & \phantom{II}-0.7&$\pm$&0.2 &      \multicolumn{3}{c}{0.}        & \multicolumn{3}{c}{0.}\\
 $f_0(980) \, \pi^+$                    &              &&&                & 16.4&$\pm$&0.7 &  4.1&$\pm$&0.6 & -3.1&$\pm$&0.2   & -4.5&$\pm$&0.3 \\
$\Kbar^*_0(1430)^0K^+$                &              &&&               &&&               &  2.4&$\pm$&0.3 & 0.48&$\pm$&0.08   & -0.7&$\pm$& 0.1\\
$f_0(1710) \, \pi^+$                     &              &&&               &&&                &&&             &  1.1 &$\pm$& 0.1 & 0.86&$\pm$&0.06\\
$f_0(1370) \, \pi^+$                     &              &&&               &&&                &&&             &&&                & 1.1 &$\pm$& 0.1 \\
\hline

\end{tabular}
\label{tab:table6}
\end{center}
\end{table*}
\begin{figure*}
\begin{center}
\includegraphics[height=9.9cm]{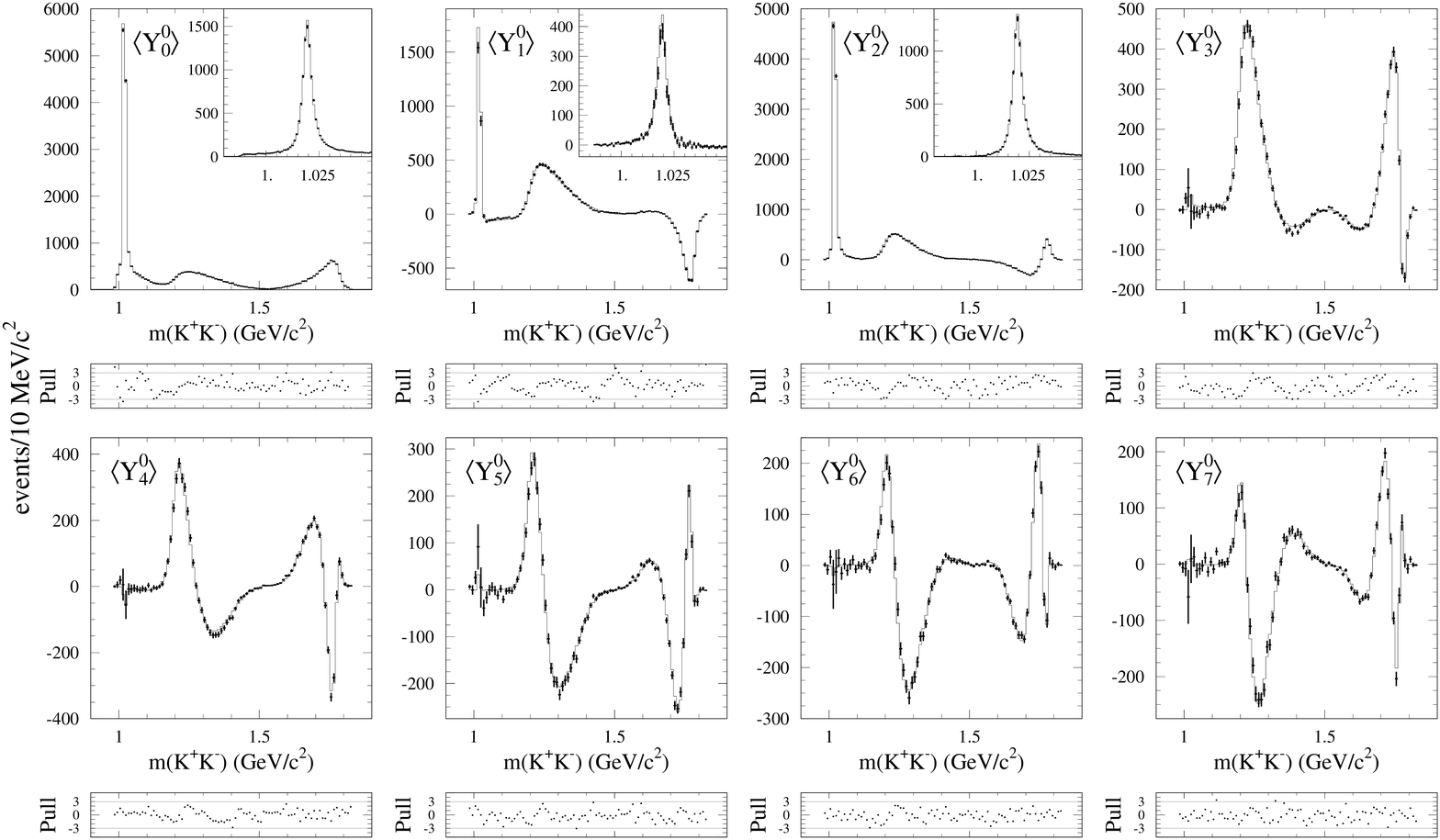}
\caption{$K^+K^-$ mass dependence of the spherical harmonic moments, $\left<Y_k^0 \right>$, obtained from the fit to the $\Ds \to \Kp\Km\pip$ Dalitz plot compared to the data moments. The data are represented by points with error
bars, the fit results by the histograms. The insets show an expanded view of the $\phi(1020)$ region.}
\label{fig:fig10}
\end{center}
\end{figure*}
\begin{figure*}
\begin{center}
\includegraphics[height=9.9cm]{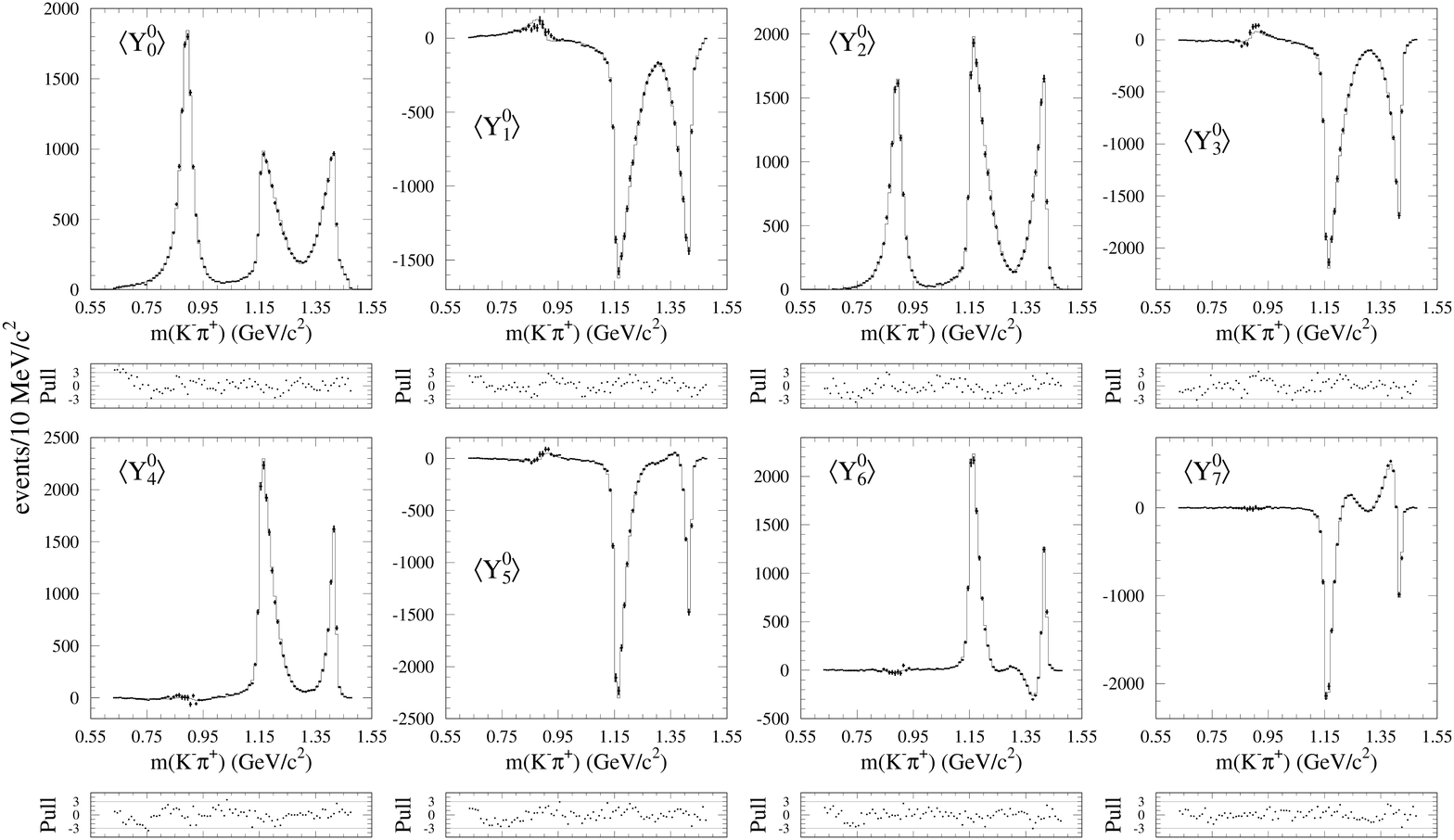}
\caption{$K^-\pi^+$ mass dependence of the spherical harmonic moments, $\left<Y_k^0 \right>$, obtained from the fit to the $\Ds \to \Kp\Km\pip$ Dalitz plot compared to the data moments. The data are represented by points with error
bars, the fit results by the histograms.}
\label{fig:fig11}
\end{center}
\end{figure*}

We notice that the width is about 3~\mev lower than that in Ref.~\cite{Nakamura:2010zzi}. 
However this measurement is consistent with results 
from other Dalitz plot analyses~\cite{:2009tr}.
\item{}
The $f_0(1370)$ contribution is also left free in the fit, and we obtain the following parameter values:
\begin{equation}
\begin{aligned}
m_{f_0(1370)}=& \, (1.22 \pm 0.01_{\rm stat} \pm 0.04_{\rm sys}) \gevcc \\
\Gamma_{f_0(1370)}=& \, (0.21 \pm 0.01_{\rm stat}  \pm 0.03_{\rm sys}) \gev
\end{aligned}
\end{equation}
These values are within the broad range of values measured by other experiments~\cite{Nakamura:2010zzi}.
\item{}
A nonresonant contribution, represented by a constant complex amplitude, was included 
in the fit function. However this contribution was found to be consistent with zero, and therefore is excluded 
from the final fit function. 
\item{} In a similar way contributions from the $K^*_1(1410)$, $f_0(1500)$, $f_2(1270)$, and $f_2'(1525)$ are found to be 
consistent with zero.
\item{} The replacement of the $K^*_0(1430)$ by the LASS parametrization~\cite{Aston:1987ir} of the 
entire $K \pi$ $\mathcal{S}$-wave does not improve the fit quality.
\item{} The fit does not require any contribution from the $\kappa(800)$~\cite{Aitala:2002kr}. 
\end{itemize}
The results of the best fit ($\chi^2/NDF=2843/(2305-14)=1.24$) are superimposed on the Dalitz plot projections in Fig.~\ref{fig:fig9}. Other recent high statistics charm Dalitz plot analyses at \babar~\cite{delAmoSanchez:2010xz} have shown that a significant contribution to the $\chi^2/NDF$ can arise from imperfections in modelling experimental effects.
The normalized fit residuals shown under each distribution (Fig.~\ref{fig:fig9}) are given by ${\rm Pull}=(N_{\rm data} - N_{\rm fit})/\sqrt{N_{\rm data}}$. The data are well reproduced in all the projections. We observe some disagreement in the $K^-\pi^+$ projection below 0.5~\gevcccc. It may be due to a poor parametrization of the background in this limited mass region. A systematic uncertainty takes such effects in account (Sec.~\ref{sec:syst}). The missing of a $K \pi$ $\mathcal{S}$-wave amplitude in the $\Km \pip$ low mass region may be also the source of  such disagreement.

Another way to test the fit quality  is to project the fit results onto the $\left<Y^0_k \right>$ moments, shown in Fig.~\ref{fig:fig10} for the $\Kp \Km$ system and Fig.~\ref{fig:fig11} for the $\Km \pip$ system. 
We observe that the fit results reproduce the data projections for moments up to $k=7$, indicating that the fit describes the details of the Dalitz plot structure very well. The $\Km \pip$ $\left<Y^0_3 \right>$ and $\left<Y^0_5 \right>$ moments show activity in the $\Kstarzbm$ region which the Dalitz plot analysis relates to interference between the $\Kstarzbm K^+$ and $f_0(1710)\pi^+$ decay amplitudes. This seems to be a reasonable explanation for the failure of the model-independent $K^-\pi^+$ analysis (Sec.~\ref{sec:kpi_swave}), although the fit still does not provide a good description of the $\left<Y^0_3 \right>$ and $\left<Y^0_5 \right>$ moments in this mass region.

We check the consistency of the Dalitz plot results and those of the analysis described in Sec.~\ref{sec:sec_pwa_b}. We compute
the amplitude and phase of the $f_0(980)$/$\mathcal{S}$-wave relative to the $\phi(1020)$/$\mathcal{P}$-wave and find  good agreement. 

\begin{table*}
\caption{
Comparison of the fitted decay fractions with the Dalitz plot analyses performed by E687 and CLEO-c collaborations.}

\begin{center}
\begin{tabular}{l r@{}c@{}l r@{}c@{}l r@{}c@{}l }
\hline
\multirow{2}{*}{Decay mode}       & \multicolumn{9}{c}{Decay fraction (\%)}\\
                                  & \multicolumn{3}{c}{\babar}   & \multicolumn{3}{c}{E687}                  &   \multicolumn{3}{c}{CLEO-c}                \\
\hline
\hline                                        
$\Kstarzbm K^+$\phantom{{\LARGE I}} & $47.9 \, \pm \,$& $0.5$ &$\, \pm \, 0.5$  & $47.8 \, \pm \,$ & $4.6$ & $\, \pm \, 4.0$   &  $47.4 \, \pm \,$ & $1.5$ & $\, \pm \,0.4$         \\ %
$\phi(1020) \, \pi^+$            &  $41.4 \, \pm \,$ & $0.8$ & $\, \pm \, 0.5$  & $39.6 \, \pm \,$& $3.3$ &$\, \pm \, 4.7$   &  $42.2 \, \pm \,$ & $1.6$ & $\, \pm \, 0.3 $      \\ %
$f_0(980) \, \pi^+$              &  $16.4 \, \pm \,$ & $0.7$ &$\, \pm \, 2.0$ & $11.0 \, \pm \, $ & $3.5$ & $\, \pm 2.6$   &  $28.2 \, \pm \,$ & $1.9$ & $\, \pm \, 1.8$      \\ %
$\Kbar^*_0(1430)^0 K^+$       &  $2.4 \, \pm \,$ & $0.3$ & $\, \pm \, 1.0$    & $9.3 \, \pm \,$& $3.2$ & $\, \pm 3.2$   &  $3.9 \, \pm \,$ & $0.5$ & $\, \pm \, 0.5$        \\ %
$f_0(1710) \, \pi^+$             &  $1.1 \, \pm \,$ &  $0.1$ & $\, \pm \, 0.1$  & $3.4 \, \pm \,$ & $2.3$ & $\, \pm 3.5$    &  $3.4 \, \pm \,$ & $0.5$ & $\, \pm \, 0.3$           \\ %
$f_0(1370) \, \pi^+$             &  $1.1 \, \pm \,$ & $0.1$ & $\, \pm \, 0.2$  & \multicolumn{3}{c}{---}                  &  $4.3 \, \pm \,$ & $0.6$ & $\, \pm \, 0.5$        \\ %
\hline
Sum                      &  $ 110.2 \, \pm \,$ & $0.6$ & $\, \pm \, 2.0$ & \multicolumn{3}{c}{111.1}                 & $129.5 \, \pm \,$ &$4.4$&$\, \pm \, 2.0$                 \\ %
$\chi^2/NDF$              &  \multicolumn{3}{c}{$\frac{2843}{(2305-14)}=1.2$}    & \multicolumn{3}{c}{$\frac{50.2}{33}=1.5$}          &  \multicolumn{3}{c}{$\frac{178}{117}=1.5$}           \\ %
Events             &  \multicolumn{3}{c}{$96307 \pm 369$}                  & \multicolumn{3}{c}{$701 \, \pm \, 36$}            &  \multicolumn{3}{c}{$12226 \, \pm \, 22$}          \\ %
\hline
\end{tabular}
\end{center}
\label{tab:tablex}
\end{table*}

\subsection{Systematic errors}
\label{sec:syst}

Systematic errors given in Table~\ref{tab:table5} and in other quoted results take into account:
\begin{itemize}
\item Variation of the $R_r$ and $R_{\Ds}$ constants in the Blatt-Weisskopf penetration factors within the range [0-3]~GeV$^{-1}$ 
and [1-5]~GeV$^{-1}$, respectively.
\item Variation of fixed resonance masses and widths within the $\pm 1\sigma$ error range quoted in Ref.~\cite{Nakamura:2010zzi}.
\item Variation of the efficiency parameters within $\pm 1\sigma$ uncertainty.
\item Variation of the purity parameters within $\pm 1\sigma$ uncertainty.
\item Fits performed with the use of the lower/upper sideband only to parametrize the background.
\item Results from fits with alternative sets of signal amplitude contributions that give equivalent Dalitz plot descriptions and similar sums of fractions.
\item Fits performed on a sample of $100,000$ events selected by applying a looser likelihood-ratio criterion 
but selecting a narrower ($\pm 1 \sigma_{\Ds}$) signal region. 
For this sample the purity is roughly the same as for the nominal sample ($\simeq 94.9\%$).
\end{itemize}

\begin{figure*}
\begin{center}
\includegraphics[width=\textwidth]{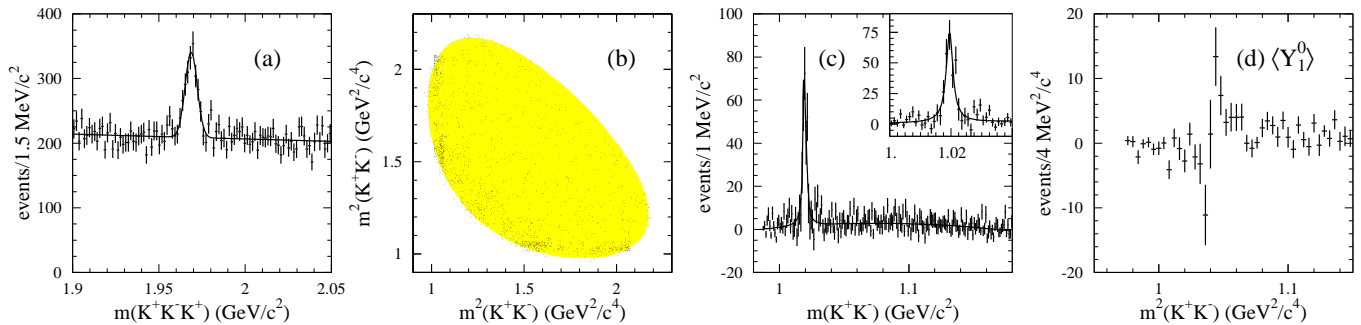}
\caption{(a) $K^+ K^- K^+$ mass spectrum showing a $\Ds$ signal. The curve is the result of the fit described in the text.
(b) Symmetrized Dalitz plot, (c) $\Kp \Km$ mass spectrum (two combinations per event), and (d) the $\left<Y^0_1\right>$ moment. The insert 
in (c) shows an expanded view
of the $\phi(1020)$ region. The Dalitz plot and its projection are background subtracted and efficiency corrected. The curve results from the fit described in the text.}
\label{fig:fig12}
\end{center}
\end{figure*}

\subsection{Comparison between Dalitz plot analyses of {\boldmath{$\protect \Ds \to \Kp \Km \pip$}}}
Table~\ref{tab:tablex} shows a comparison of the Dalitz plot fit fractions, shown in Table~\ref{tab:table5}, with the results of the analyses performed by the E687~\cite{Frabetti:1995sg} and CLEO~\cite{:2009tr} collaborations. The E687 model is improved by adding a $f_0(1370)$ amplitude and leaving the $\Kstarzbm$ parameters free in the fit. We find that the $\Kstarzbm$ width (Eq.~\ref{eq:m_g_892}) is about 3~\mev lower than that in Ref.~\cite{Nakamura:2010zzi}. This result is consistent with the width measured by CLEO-c collaboration ($\Gamma_{\Kstarzbm} = 45.7 \pm 1.1 \mev$). 

What is new in this analysis is the parametrization of the $\Kp\Km$ $\mathcal{S}$-wave at the $\Kp\Km$ threshold. While E687 and CLEO-c used a coupled channel BW (Flatt\'e) amplitude~\cite{Flatte:1972rz} to parametrize the $f_0(980)$ resonance, we use the model independent parametrization described in Section~\ref{sec:sec_pwa_b}. This approach overcomes the uncertainties that affect the coupling constants $g_{\pi\pi}$ and $g_{KK}$ of the $f_0(980)$, and any argument about the presence of an $a(980)$ meson decaying to $\Kp\Km$. The model, described in this paper, returns a more accurate description of the event distribution on the Dalitz plot ($\chi^2/\nu=1.2$) and smaller $f_0(980)$ and total fit fractions respect to the CLEO-c result.  In addition the goodness of fit in this analysis is tested by an adaptive binning $\chi^2$, a tool more suitable when most of the events are gathered in a limited region of the Dalitz plot.

Finally we observe that the phase of the $\phi(1020)$ amplitude ($166^\circ \pm 1^\circ \pm 2^\circ$) is consistent with the E687 result ($178^\circ\pm20^\circ\pm24^\circ$) but is roughly shifted by $180^\circ$ respect to the CLEO-c result ($-8^\circ \pm 4^\circ \pm 4^\circ$).

\section{Singly-Cabibbo-Suppressed {\boldmath$\protect\Ds \to \Kp \Km \Kp$}, and Doubly-Cabibbo-Suppressed {\boldmath$\protect\Ds \to \Kp \Kp \pim$} decay}
\label{sec:sec_BR}
In this section we measure the branching ratio of the SCS decay channel (\ref{eq:eq2}) 
and of the DCS decay channel (\ref{eq:eq3}) with respect to the CF decay channel (\ref{eq:eq1}).
The two channels are reconstructed using the method described in Sec.~\ref{sec:sec_ev_sel} 
with some differences related to the particle identification of the $\Ds$ daughters.
For channel (\ref{eq:eq2}) we require the identification 
of three charged kaons while for channel (\ref{eq:eq3}) we require the identification of one 
pion and two kaons having the 
same charge. We use both the $\Dss$ identification and the likelihood-ratio
 to enhance signal with respect to background as described in Sec.~\ref{sec:sec_ev_sel}. 

The ratios of branching fractions are computed as:
\begin{equation}
\frac{\BR(\Ds \to \Kp \Km \Kp)}{\BR(\Ds \to \Kp \Km \pip)} \kern-0.3em = \kern-0.3em \frac{N_{\Ds \to \Kp \Km \Kp}}{N_{\Ds \to \Kp \Km \pip}}\kern-0.3em \times \kern-0.3em \frac{\epsilon_{\Ds \to \Kp \Km \pip}}{\epsilon_{\Ds \to \Kp \Km \Kp}}
\end{equation}
and
\begin{equation}
\frac{ \BR(\Ds \to K^+ K^+ \pim)}{ \BR(\Ds \to K^+ K^- \pi^+)} \kern-0.3em = \kern-0.3em \frac{N_{\Ds \to K^+ K^+ \pim}}{N_{\Ds \to K^+ K^- \pi^+}} \times \frac{\epsilon_{\Ds \to K^+ K^- \pi^+}}{\epsilon_{\Ds \to K^+ K^+ \pim}}.
\end{equation}
\noindent
Here the $N$ values represent the number of signal events for each channel, and the $\epsilon$ values indicate 
the corresponding detection efficiencies.

To compute these efficiencies, we generate signal MC samples having uniform distributions across the Dalitz plots.
These MC events are reconstructed as for data events, and the same particle-identification criteria are applied. 
Each track is weighted
by the data-MC discrepancy in particle identification efficiency obtained 
independently from high statistics control samples. A systematic uncertainty is assigned to the use of this weight.
The generated and reconstructed Dalitz plots are divided into $50 \times 50$ cells and the Dalitz plot efficiency
is obtained as the ratio of reconstructed to generated content of each cell. In this way the
efficiency for each event depends on its location on the Dalitz plot.
By varying the likelihood-ratio criterion, the  
sensitivity $S$ of $\Ds \to K^+ K^- K^+$ is maximized. The sensitivity is defined as $S = N_s/\sqrt{N_s + N_b}$, where $s$ and $b$ indicate signal and background. To reduce systematic uncertainties, we then apply the same likelihood-ratio criterion to the $\Ds \to K^+ K^- \pi^+$ decay.
We then repeat this procedure to find an independently optimized selection criterion for the $\Ds \to K^+ K^+ \pim$ to $\Ds \to K^+ K^- \pi^+$ ratio.

The branching ratio measurements are validated
using a fully inclusive $e^+e^-\to c \bar c$ MC simulation
incorporating all known charmed meson decay modes. The MC
 events are subjected to the same reconstruction,
event selection, and analysis procedures as for the data.
The results are found to be consistent, within statistical
uncertainty, with the branching fraction values used in the
MC generation.

\subsection{Study of {\boldmath$\protect\Ds \to K^+ K^- K^+$}}
\label{sec:sec_kkk}
The resulting $K^+ K^- K^+$ mass spectrum is shown in Fig.~\ref{fig:fig12}(a). The $\Ds$ yield is obtained by fitting the mass spectrum using a Gaussian function for the signal, and 
a linear function for the background. The resulting yield is reported in Table~\ref{tab:table1}.

The systematic uncertainties are summarized in Table~\ref{tab:table7} and are evaluated as follows:
\begin{itemize}
\item{} The effect of MC statistics is evaluated by randomizing each efficiency cell on the Dalitz plot
according to its statistical uncertainty.
\item{} The selection made on the $\Dss$ candidate $\Delta m$ is varied to $\pm$2.5$\sigma_{\Dss}$ and $\pm$1.5$\sigma_{\Dss}$.
\item{} For particle identification we make use of high statistics control samples to assign 1\% uncertainty to each kaon and 0.5\% to each pion. 
\item{} The effect of the likelihood-ratio criterion is studied by measuring the branching ratio for different choices.
\end{itemize}
\begin{table}[!htbp]
\caption{Summary of systematic uncertainties on the measurement of the $\Ds \to K^+ K^- K^+$ branching ratio.}
\begin{center}
\begin{tabular}{cc} \hline
 Uncertainty & $\frac{ \BR(\Ds \to K^+ K^- K^+)}{ \BR(\Ds \to K^+ K^- \pi^+)}$ \\ \hline \hline
 MC statistics                   &  2.6 \%  \\ 
 $\Delta m$                      &  0.3 \%  \\ 
 Likelihood-ratio                &  3.5 \%  \\ 
 PID                             &  1.5  \%  \\ \hline 
 Total                           &  4.6 \%  \\ \hline
\end{tabular}
\label{tab:table7}
\end{center}
\end{table}

We measure the following branching ratio:
\begin{equation}
\frac{ \BR(\Ds \to K^+ K^- K^+)}{ \BR(\Ds \to K^+ K^- \pi^+)} = (4.0 \pm 0.3_{\rm stat} \pm 0.2_{\rm syst}) \times 10^{-3}.
\end{equation}
\noindent

\begin{figure*}
\begin{center}
\includegraphics[width=0.951\textwidth]{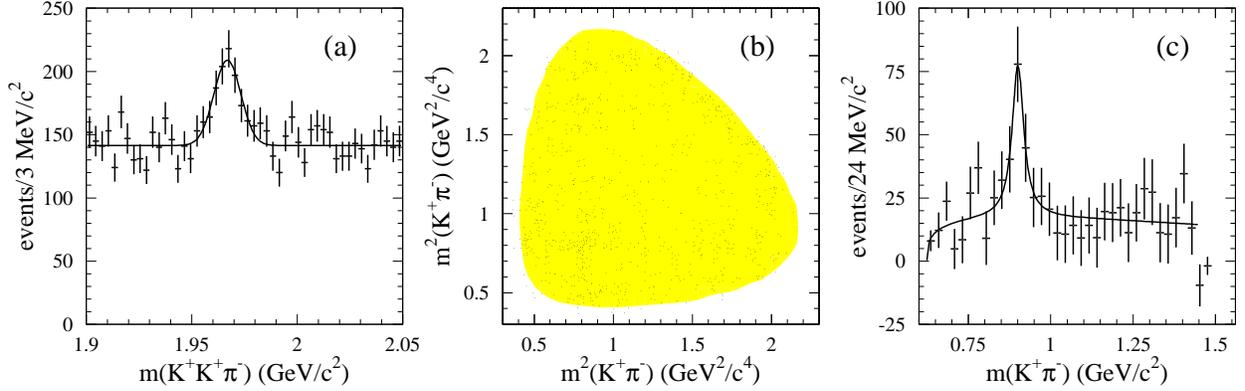}
\caption{(a) $K^+ K^+ \pi^-$ mass spectrum showing a $\Ds$ signal. (b) Symmetrized Dalitz plot for $\Ds \to K^+ K^+ \pi^-$ decay. (c) $K^+ \pi^-$ mass distribution (two combinations per event). The Dalitz plot and its projection are background subtracted and efficiency corrected. The curves result from  the fits described in the text.}
\label{fig:fig13}
\end{center}
\end{figure*}

A Dalitz plot analysis in the presence of a high level of background is difficult, therefore we can only extract empirically some information on the decay. Since there are two identical kaons into the final state, the Dalitz plot is symmetrized by plotting two combinations per event ($[m^2(K^-K^+_1), m^2(K^-K^+_2)]$ and $[m^2(K^-K^+_2), m^2(K^-K^+_1)]$).
The symmetrized Dalitz plot in the $\Ds \to K^+ K^- K^+$ signal region, corrected for efficiency and background-subtracted, is shown in Fig.~\ref{fig:fig12}(b).
It shows two bands due to the $\phi(1020)$ and no other structure, indicating a large contribution via $\Ds \to \phi(1020)K^+$. To test the
possible presence of $f_0(980)$, we plot, in Fig.~\ref{fig:fig12}(d), the distribution of the $\left<Y^0_1\right>$ moment; $\theta$ is the angle between the $K^+$ direction in the $K^+ K^-$ rest frame and the prior direction of the $K^+ K^-$
 system in the $D^+_s$ rest frame.
We observe the mass dependence characteristic of interference between $\mathcal{S}$- and $\mathcal{P}$-wave amplitudes, and conclude that there is a contribution from $\Ds \to f_0(980)K^+$ decay, although its branching fraction cannot be determined in the present analysis.

An estimate of the $\phi(1020)K^+$ fraction can be obtained from a fit to the $K^+K^-$ mass
 distribution (Fig.~\ref{fig:fig12}(c)). The mass spectrum is fitted using a relativistic BW for the
$\phi(1020)$ signal, and a second order polynomial for the background. We obtain:
\begin{eqnarray}
\frac{\BR(\Ds \to \phi K^+)\cdot\BR(\phi \to K^+ K^-)}{\BR(\Ds \to K^+ K^- K^+ )} & = & \nonumber \\
0.41  \pm  0.08\kern-0.6em&_{\rm stat}&\kern-0.5em \pm 0.03_{\rm syst}.
\end{eqnarray}

The systematic uncertainty includes the contribution due to $\Delta m$ and the likelihood-ratio criteria, the fit model, and the background parametrization.

\subsection{Study of {\boldmath$\Ds \to K^+K^+\pi^-$}}

Figure~\ref{fig:fig13}(a) shows the $K^+K^+\pi^-$ mass spectrum.
A fit with a Gaussian signal function and a linear background function gives the yield presented in Table~\ref{tab:table1}.
To minimize systematic uncertainty, we apply the same likelihood-ratio criteria to the $K^+ K^+ \pi^-$ and $K^+ K^- \pi^+$ final states,  
and correct for the efficiency evaluated on the Dalitz plot. 
The branching ratio which results is:
\begin{equation}
\frac{ \BR(\Ds \to K^+ K^+ \pi^-)}{ \BR(\Ds \to K^+ K^- \pi^+)} = (2.3 \pm 0.3_{\rm stat} \pm 0.2_{\rm syst}) \times 10^{-3}.
\end{equation}
This value is in good agreement with the Belle measurement: $\frac{ \BR(\Ds \to K^+ K^+ \pi^-)}{ \BR(\Ds \to K^+ K^- \pi^+)} =(2.29 \pm 0.28 \pm 0.12) \times 10^{-3}$~\cite{Ko:2009tc}.

Table~\ref{tab:table8} lists the results of the systematic studies performed
for this measurement; these are similar to those used in Sec.~\ref{sec:sec_kkk}. The particle identification systematic is not taken in account because the final states differ only in the charge assignments of the daughter tracks.
\begin{table}[!htbp]
\caption{Summary of systematic uncertainties in the measurement of the $\Ds \to K^+ K^+ \pi^-$ relative branching fraction.}
\begin{center}
\begin{tabular}{cc} \hline
 Uncertainty & $\frac{ \BR(\Ds \to K^+ K^+ \pi^-)}{ \BR(\Ds \to K^+ K^- \pi^+)}$ \\ \hline \hline
 MC statistics                   &  0.04 \%  \\ 
 $\Delta m$                      &  4.7 \%  \\ 
 Likelihood-ratio                &  6.0 \%  \\ \hline 
 Total                           &  7.7 \%  \\ \hline
\end{tabular}
\label{tab:table8}
\end{center}
\end{table}

The symmetrized Dalitz plot for the signal region, corrected for efficiency and  background-subtracted, is shown in Fig.~\ref{fig:fig13}(b). 
We observe the presence of a significant $\Kstarzm$ signal, which is more evident in the $\Kp \pim$ mass distribution, shown in Fig.~\ref{fig:fig13}(c).
Fitting this distribution using a relativistic $\mathcal{P}$-wave BW signal function and a threshold function, 
we obtain the following fraction for this contribution.
\begin{eqnarray}
\frac{\BR(\Ds \to \Kstarzm K^+)\cdot \BR(\Kstarzm \to K^+ \pi^-)}{\BR(\Ds \to K^+ K^+ \pi^- )} = \nonumber \\
0.47 \pm 0.22_{\rm stat} \pm 0.15_{\rm syst}.
\end{eqnarray}
Systematic uncertainty contributions include those from $\Delta m$ and the likelihood-ratio criteria, the fitting model, and the background parametrization. 

The symmetrized Dalitz plot shows also an excess of events at low $\Kp\Kp$ mass, which may be due to a Bose-Einstein correlation effect~\cite{Goldhaber:1960sf}. We remark, however, that this effect is not visible in $\Ds \to \Kp\Km\Kp$ decay (Fig.~\ref{fig:fig12}(b)).

\section{Conclusions}
\label{sec:sec_sum}
In this paper we perform a high statistics Dalitz plot analysis of $\Ds \to \Kp \Km \pip$, and extract amplitudes
and phases for each resonance contributing to this decay mode. We also make a new measurement of the 
$\mathcal{P}$-wave/$\mathcal{S}$-wave ratio in the $\phi(1020)$ region. The $\Kp \Km$ $\mathcal{S}$-wave is extracted in
a quasi-model-independent way, and complements the $\pip \pim$ $\mathcal{S}$-wave measured by this experiment
in a previous publication~\cite{:2008tm}. Both measurements can be used to obtain new information on the properties of the
$f_0(980)$ state~\cite{Pennington:2007zy}. We also measure  the relative and partial branching fractions for the SCS $\Ds \to K^+ K^- K^+$ and DCS $\Ds \to K^+ K^+ \pi^-$ decays with high precision.

\section{Acknowledgments}
We are grateful for the 
extraordinary contributions of our \pep2\ colleagues in
achieving the excellent luminosity and machine conditions
that have made this work possible.
The success of this project also relies critically on the 
expertise and dedication of the computing organizations that 
support \babar.
The collaborating institutions wish to thank 
SLAC for its support and the kind hospitality extended to them. 
This work is supported by the
US Department of Energy
and National Science Foundation, the
Natural Sciences and Engineering Research Council (Canada),
the Commissariat \`a l'Energie Atomique and
Institut National de Physique Nucl\'eaire et de Physique des Particules
(France), the
Bundesministerium f\"ur Bildung und Forschung and
Deutsche Forschungsgemeinschaft
(Germany), the
Istituto Nazionale di Fisica Nucleare (Italy),
the Foundation for Fundamental Research on Matter (The Netherlands),
the Research Council of Norway, the
Ministry of Education and Science of the Russian Federation, 
Ministerio de Ciencia e Innovaci\'on (Spain), and the
Science and Technology Facilities Council (United Kingdom).
Individuals have received support from 
the Marie-Curie IEF program (European Union), the A. P. Sloan Foundation (USA) 
and the Binational Science Foundation (USA-Israel).

\end{document}